\documentclass[aps,prb,reprint,showkeys]{revtex4-2}

\usepackage{graphicx}
\usepackage{soul,color}
\sethlcolor{green}

\begin{document}

\title{Piezoelectric properties of ferroelectric perovskite superlattices with polar discontinuity}

\author{Alexander I. Lebedev}
\email[]{swan@scon155.phys.msu.ru}
\affiliation{Physics Department, Moscow State University, 119991 Moscow, Russia}

\date{\today}

\begin{abstract}
The stability of a high-symmetry $P4mm$~polar phase in seventeen short-period
ferroelectric perovskite superlattices with polar discontinuity is studied from
first principles within the density-functional theory. It is shown that in most
superlattices this phase exhibits either the ferroelectric instability or the
antiferrodistortive one, or both of them. For each superlattice, the ground-state
structure, the structure of possible metastable phases, the spontaneous polarization,
and piezoelectric properties are calculated. The properties of superlattices with
the polar discontinuity are compared with those of superlattices with broken
symmetry and of ordinary superlattices, which have no the polar discontinuity.
It is shown that high piezoelectric coefficients (up to 150--270~pC/N) in some
superlattices with the polar discontinuity are due to the appearance of strong
lattice distortions, whose symmetry follows that of a low-lying polar phonon mode
of the ground-state structure, under the influence of external strain.

\texttt{DOI: 10.1016/j.commatsci.2020.110113}
\end{abstract}


\maketitle

\section{Introduction}

In recent years, much attention has been paid to studies of low-dimensional structures
in which new physical phenomena that have no analogues in bulk materials have been
discovered. Due to these new functional properties, these materials are considered
as very promising for their use in electronics. Ferroelectric superlattices
(SLs)---quasi-two-dimensional artificial periodic structures whose properties can
be tailored to obtain the necessary functionality---belong to this interesting
class of materials.

Most of the earlier experimental~\cite{JApplPhys.72.2840,JapJApplPhys.33.5192,
ApplPhysLett.65.1970,ApplPhysLett.70.321,ApplPhysLett.72.1394,PhysRevLett.80.4317,
ApplPhysLett.77.1520,ApplPhysLett.77.3257,JapplPhys.91.2290,PhysRevLett.88.097601,
ApplPhysLett.80.3581,ApplPhysLett.82.2118,JApplPhys.93.1180,PhysRevLett.95.177601,
ApplPhysLett.89.092905,Science.313.1614,AdvMater.19.4153,PhysRevLett.104.187601,
PhysSolidState.54.1026,PhysRevLett.109.167601,PhysSolidState.57.486,Nature.568.368,
SciRep.9.18948} and theoretical~\cite{PhysRevB.59.12771,PhysRevLett.84.5636,
ApplPhysLett.82.1586,PhysRevB.69.184101,Nature.433.395,PhysRevB.71.100103,
ApplPhysLett.87.052903,ApplPhysLett.87.102906,HandbookChap134,PhysRevB.76.020102,
ApplPhysLett.91.112914,Nature.452.732,PhysRevLett.101.087601,PhysRevB.79.024101,
PhysSolidState.51.2324,PhysSolidState.52.1448,PhysRevB.83.020104,PhysRevLett.107.217601,
PhysStatusSolidiB.249.789,PhysRevB.85.184105,PhysSolidState.55.1198,
ComputMaterSci.91.310,Nature.534.524,NanoLett.17.2246,ActaMater.152.155}
studies of ferroelectric superlattices with the perovskite
structure were performed on II-IV/II-IV or I-V/I-V type superlattices (here, the
numbers indicate the valence of atoms entering the $A$ and $B$~sites of the $AB$O$_3$
perovskite structure). In such superlattices, there are no double electric layers
at the interface between two dielectric materials, and therefore both the bulk
and interfaces of these superlattices remain macroscopically electrically neutral.

Studies of the SrTiO$_3$/LaAlO$_3$ heterostructures have revealed new interesting
phenomena that appear in these structures as a result of so-called polar
discontinuity---of a polarization jump induced by effective $\pm e/2$ charges
per unit cell which appear at the interface between SrTiO$_3$ and LaAlO$_3$
because the charged character of LaO$^+$ and AlO$_2^-$ layers~\cite{Nature.427.423}.
These phenomena include the appearance of a conducting layer near the interface
(a two-dimensional electron gas), its magnetism, and even
superconductivity~\cite{Nature.427.423, NatureMater.6.493, Science.317.1196}. These
phenomena can be controlled using an external electric field~\cite{Nature.456.624}.
The divergence of the electrostatic potential in such heterostructures, which
can result in the appearance of conductive layers at the interface between two
dielectrics, is called a polar catastrophe. The possibility of the appearance of
the two-dimensional electron gas at the interface between a ferroelectric and a
nonpolar dielectric in perovskites was systematically studied in \cite{SciRep.6.34667}.
Later it was realized that conductive layers can also be obtained in ferroelectric
structures without the polar discontinuity~\cite{PhysRevB.92.115406}. This made
it possible to create new types of electronic devices---ferroelectric structures
with switchable giant tunneling conductivity \cite{ApplPhysLett.107.232902,
PhysRevB.94.155420}; the idea of such devices was proposed earlier in
\cite{Science.313.181, NanoLett.9.427}. Note that the appearance of similar
phenomena can also be anticipated in epitaxial films of II-IV and I-V perovskites,
which are grown on DyScO$_3$, GdScO$_3$, NdScO$_3$, NdGaO$_3$, and LaGaO$_3$
substrates to create a biaxial strain in them, since the polar discontinuity
is possible in these structures~\cite{ChinPhysB.28.047101}.

Until now, theoretical studies of superlattices with the polar discontinuity have
focused on studying the distribution of polarization and electric field in these
structures and on finding the conditions for the appearance of a two-dimensional
electron gas at the interface~\cite{PhysRevB.79.100102,PhysRevLett.103.016804,
PhysRevB.80.045425,PhysRevB.80.165130,PhysRevB.80.241103,PhysRevB.81.235112,
PhysRevB.85.235109,PhysRevB.87.085305,SciRep.8.467,RCSAdv.9.35499,ChinPhysB.28.047101,
PhysChemChemPhys.21.8046}. The questions of the stability of the high-symmetry
polar structure in such SLs, possible phase transitions in them, and the physical
properties of low-symmetry phases were not analyzed. At the same time, the
ferroelectric and antiferrodistortive (AFD) instabilities characteristic of many
perovskites can result in strong distortions of the structure of SLs. This means
that the earlier predictions of their physical properties obtained without taking
these distortions into account may be incorrect.

One of the questions that remains insufficiently studied is the question about
the piezoelectric properties of superlattices with the polar discontinuity. The
calculations of these properties are limited to the calculations for the
high-symmetry tetragonal $P4mm$~phase of such SLs~\cite{JApplPhys.109.066107,
ChinPhysB.27.027701}. It is known that the record-high piezoelectric properties
of PbTiO$_3$-based solid solutions near the morphotropic boundary~\cite{JApplPhys.82.1804}
are due to the ease of inclination of the polarization vector from the
[100] direction to the [111] one under the influence of the electric
field~\cite{PhysRevLett.84.5423} or mechanical stress. The polar discontinuity
in SLs allows, under certain conditions, to obtain high values of irreversible
polarization in their structures. This is why it was interesting to check
whether it is possible to get high piezoelectric coefficients in superlattices
with the polar discontinuity using the strain-induced inclination of the
polarization vector. In Ref.~\cite{JApplPhys.109.066107}, SLs with the polar
discontinuity were already considered as a way to obtain stable, weakly
temperature-dependent piezoelectric properties.

\section{Calculation technique}

\begin{figure}
\centering
\includegraphics{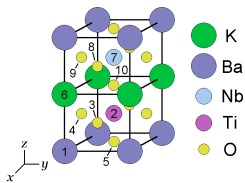}
\caption{Geometry of a typical superlattice.}
\label{fig1}
\end{figure}

In this work, the properties of seventeen free-standing [001]-oriented
short-period II-IV/I-V, II-IV/III-III, and I-V/III-III superlattices with the
polar discontinuity and a thickness of individual layers of one unit cell are
studied from first principles. For comparison, similar calculations are
performed for several superlattices with broken symmetry. The latter SLs have
no the polar discontinuity, but their high-symmetry polar structure is formed
by a sequence of layers with no mirror symmetry. In addition, the obtained
results are compared with the results for ordinary superlattices with the
mirror symmetry of the layer sequence. The geometry of the superlattices is
shown in Fig.~\ref{fig1}.

The choice of free-standing superlattices for our study stems from the fact that,
when any coherently strained short-period superlattice is grown on a substrate,
a transition layer containing misfit dislocations and other defects with a
thickness of $\sim$100~{\AA} appears at the interface as a result of a mismatch
between the equilibrium in-plane lattice parameter of the SL and that of the
substrate. In the defect-free part of the superlattice, the in-plane lattice
parameter relaxes to the lattice parameter of the free-standing superlattice.
This is why in the majority of rather thick experimentally grown superlattices,
the in-plane lattice parameter is close to that of free-standing superlattices.

First-principles calculations were performed within the density functional theory
using the \texttt{ABINIT} program~\cite{abinit} and norm-conserving pseudopotentials
constructed according to the RRKJ scheme~\cite{PhysRevB.41.1227} in the local
density approximation (LDA), like in~\cite{PhysSolidState.51.362}. The cutoff
energy was 30~Ha (816~eV) with the exception of Ta-containing systems in which
it was 40~Ha (1088~eV). Integration over the Brillouin zone was carried out using
a 8$\times$8$\times$4 Monkhorst--Pack mesh for high-symmetry structures and
meshes with equivalent density of ${\bf k}$-points for low-symmetry phases.
The equilibrium lattice parameters and atomic positions were calculated by
relaxing forces acting on the atoms to values less than $2 \cdot 10^{-6}$~Ha/Bohr
(0.1~meV/{\AA}). The phonon spectra, the tensors of piezoelectric stress
coefficients $e_{i\mu}$, and the elastic compliance tensors $S_{\mu\nu}$ were
calculated using the density-functional perturbation theory. The $e_{i \mu}$
values were then converted to the piezoelectric strain coefficients $d_{i\nu}$
using the formula $d_{i\nu} = e_{i\mu} S_{\mu\nu}$, and for monoclinic structures
the tensor components were transformed to the standard setting of the monoclinic
unit cell, in which the polarization vector lies in the $xz$~plane.

\section{Results and discussion}

\subsubsection{Search for the ground-state structure}

It is known that the electrical conductivity of materials is usually a factor
that prevents the practical use of their ferroelectric properties. Since the
experiments on structures with the polar discontinuity often revealed metallic
conductivity at the interface, it was necessary first to ensure that the
superlattices under study are insulating. The calculations confirmed that
in all superlattices studied in the work, the conduction band is separated
from the valence band by a sufficiently large energy gap (see Table~S1
in the Supplementary data), and therefore all studied SLs are dielectrics.

Since the layer sequence in considered superlattices does not admit the reversal
of $z \to -z$ (Fig.~\ref{fig1}), the superlattices are always polar and their
high-symmetry phase is described by the $P4mm$~space group. However, this
structure can exhibit various instabilities characteristic of crystals with
the perovskite structure: either the ferroelectric instability or the
antiferrodistortive one, or both of them simultaneously. In this work, the
ground state of
superlattices was searched in the traditional way~\cite{PhysSolidState.51.2324,
PhysSolidState.52.1448}. First, the structures resulting from the condensation
of all unstable phonons found in the phonon spectrum of the $P4mm$~phase were
calculated taking into account a possible degeneracy of phonon modes. Then, by
calculating the phonon frequencies at all high-symmetry points of the Brillouin
zone and the elastic tensor for these distorted structures, the stability of
the obtained solutions was checked. In the case when an instability was found
in any of these structures, the search for the ground state was continued until
a structure whose phonon spectrum has no unstable modes and whose matrix composed
of the elastic tensor components in the Voigt notation is positive definite is
found. In this case, the conclusion that a \emph{stable} phase which can be a
possible ground-state structure can be made. The problem here is that in chains
of phases generated by different octahedra rotations, several stable states can
be found, as was recently demonstrated for SrTiO$_3$~\cite{PhysSolidState.58.300}.
In this case, the ground state is the stable phase with the lowest total energy;
other stable phases should be considered as metastable. If the energy of such
metastable solutions differs little from the energy of the ground state, then
these solutions should be considered as phases that can be observed
in experiment. For them, as well as for the ground state, an analysis of their
physical properties should also be carried out.

\begin{figure}
\centering
\includegraphics{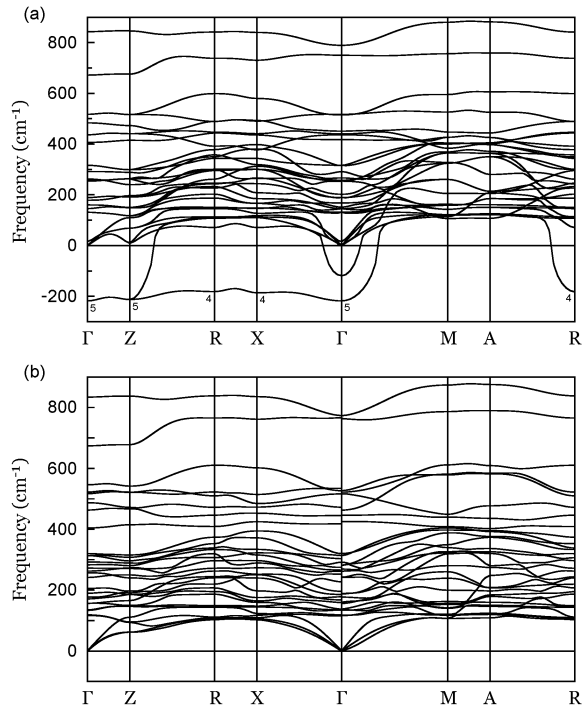}
\caption{\label{fig2}Phonon spectrum of the KNbO$_3$/BaTiO$_3$ superlattice
(a) in the high-symmetry $P4mm$~phase and (b) in the ground-state $Cm$~phase.
The numbers near the curves indicate the symmetry of unstable modes.}
\end{figure}

Calculations of the phonon spectra show that in most studied superlattices
with the polar discontinuity (except for BaTiO$_3$/LaAlO$_3$, SrTiO$_3$/LaAlO$_3$,
PbTiO$_3$/KTaO$_3$, PbTiO$_3$/LaAlO$_3$, and KTaO$_3$/LaAlO$_3$) the high-symmetry
$P4mm$~phase exhibits the ferroelectric instability with respect to the in-plane
distortion of the structure or, in other words, to the inclination of the polarization
vector. The phonon spectrum of one of such superlattices, KNbO$_3$/BaTiO$_3$, is
shown in Fig.~\ref{fig2}. It is seen that in addition to the instability at the
$\Gamma$ point, the instabilities also appear at the $X$, $R$, and $Z$~points of
the Brillouin zone. We have already encountered a similar situation in
KNbO$_3$/KTaO$_3$~\cite{PhysStatusSolidiB.249.789} and
BaTiO$_3$/BaZrO$_3$~\cite{PhysSolidState.55.1198} SLs.

\begin{figure}
\centering
\includegraphics{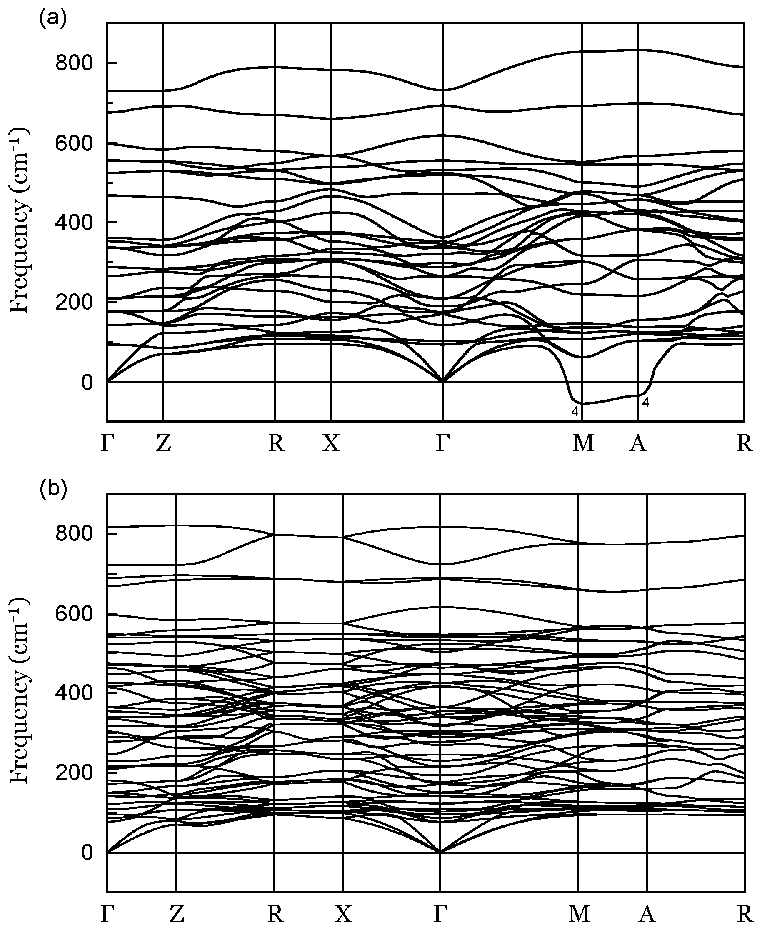}
\caption{\label{fig3}Phonon spectrum of the BaTiO$_3$/LaAlO$_3$ superlattice
(a) in the high-symmetry $P4mm$~phase and (b) in the ground-state $P4bm$~phase.
The numbers near the curves indicate the symmetry of unstable modes.}
\end{figure}

\begin{figure}
\centering
\includegraphics{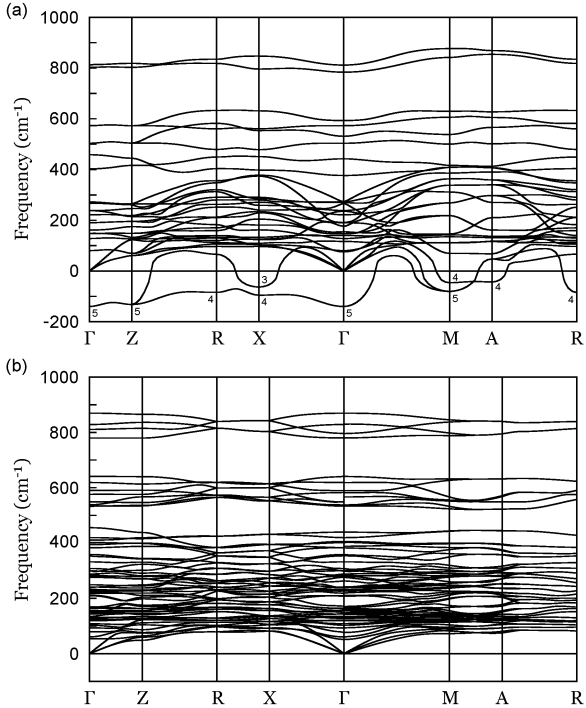}
\caption{\label{fig4}Phonon spectrum of the KNbO$_3$/SrZrO$_3$ superlattice
(a) in the high-symmetry $P4mm$~phase and (b) in the ground-state $Pc$~phase.
The numbers near the curves indicate the symmetry of unstable modes.}
\end{figure}

An instability zone, which is observed as a band of imaginary phonon frequencies
on the $\Gamma$--$Z$--$R$--$X$--$\Gamma$ line (imaginary frequencies are
represented in the figure by negative values), is a consequence of the
ferroelectric instability in the ...--Ti--O--... chains propagating in the plane
of the SL---the so-called chain instability~\cite{PhysRevLett.74.4067}. Indeed,
an analysis of the eigenvectors of these modes shows that at all the
above-mentioned points of the Brillouin zone the out-of-phase, transverse,
$xy$-polarized displacements of Ti and O atoms in chains propagating in the [100]
and [010] directions dominate in the atomic motion. In
Ba(Sr)TiO$_3$/KNb(Ta)O$_3$ superlattices, the displacements of the Nb(Ta)
atoms are small, whereas in superlattices containing no Ti atoms as well as
in the PbTiO$_3$/KNbO$_3$ SL the chain instability is observed mainly in
...--Nb--O--... chains. In BiScO$_3$/LaAlO$_3$ and PbTiO$_3$/BiScO$_3$
superlattices, additional out-of-phase, $xy$-polarized displacements of Bi and
oxygen atoms lying in the same $xy$~plane are stronger; this displacement
pattern is characteristic of BiScO$_3$. A similar displacement pattern is also
typical for unstable modes at the $Z$~point in these superlattices. At the
$\Gamma$~point, the described displacement pattern corresponds to a doubly
degenerate ferroelectric $E$~mode ($\Gamma_5$).%
    \footnote{When numbering irreducible representations in this paper, we follow
    their classification accepted at Bilbao Crystallographic Server~\cite{Bilbao}.}

\begin{table*}
\caption{\label{table1}Energies (per 10-atom supercell) of different low-symmetry
phases resulting from the condensation of unstable phonons in the high-symmetry
$P4mm$~phase of short-period KNbO$_3$/BaTiO$_3$ and KNbO$_3$/SrZrO$_3$ superlattices.}
\begin{ruledtabular}
\begin{tabular}{cccccccc}
Phase    & Unstable     & Energy   & Phase    & Unstable               & Energy \\
         & phonon       & (meV)    &          & phonon                 & (meV) \\
\hline
\multicolumn{6}{c}{KNbO$_3$/BaTiO$_3$ superlattice} \\
$P4mm$   & ---          & 0        & $Pm$     & $\Gamma_5~(\eta,0)$    & $-$27.8 \\
$Abm2$   & $R_4$        & $-$13.8  & $Cmc2_1$ & $Z_5~(\eta,\eta)$      & $-$31.0 \\
$Pma2$   & $X_4$        & $-$14.8  & $Cm$     & $\Gamma_5~(\eta,\eta)$ & $-$38.6 \\
$Pmc2_1$ & $Z_5~(\eta,0)$ & $-$21.3 & \\
\multicolumn{6}{c}{KNbO$_3$/SrZrO$_3$ superlattice} \\
$P4mm$   & ---                     & 0        & $Pc$     & $Z_5(\eta,0)+\Gamma_5(0,\xi)$ & $-$19.2 \\
$I4cm$   & $A_4$                   & $-$1.3   & $Pma2$   & $M_5~(\eta,0)$          & $-$19.9 \\
$P4bm$   & $M_4$                   & $-$2.1   & $Cm$     & $\Gamma_5~(\eta,\eta)$  & $-$20.0 \\
$Abm2$   & $R_4$                   & $-$4.7   & $P2$     & $X_3+X_4$               & $-$20.7 \\
$Pma2$   & $X_4$                   & $-$6.7   & $Cc$     & $A_4+\Gamma_5~(\eta,\eta)$ & $-$24.1 \\
$Pmm2$   & $X_3$                   & $-$12.4  & $Pm$     & $X_3+\Gamma_5~(0,\eta)$ & $-$24.9 \\
$Pmc2_1$ & $Z_5~(\eta,0)$          & $-$12.9  & $Cmm2$   & $M_5~(\eta,\eta)$       & $-$26.9 \\
$Pm$     & $\Gamma_5~(\eta,0)$     & $-$13.7  & $Pm$     & $X_3+\Gamma_5~(\eta,0)$ & $-$27.6 \\
$Pc$     & $R_4+\Gamma_5~(\eta,0)$ & $-$15.4  & $Pm$     & $M_5~(\eta,0)+\Gamma_5~(\xi,\xi)$ & $-$36.9 \\
$Cm$     & $A_4+\Gamma_5(\eta,0)$  & $-$16.0  & $Cm$     & $M_4+\Gamma_5~(\eta,0)$ & $-$39.4 \\
$Pc$     & $X_4+\Gamma_5~(\eta,0)$ & $-$17.0  & $Cm$     & $M_5~(\eta,\eta)+\Gamma_5~(0,\xi)$ & $-$40.2 \\
$Cmc2_1$ & $Z_5~(\eta,\eta)$       & $-$18.9  & $Pc$     & $M_4+\Gamma_5~(\eta,\eta)$ & $-$51.9 \\
\end{tabular}
\end{ruledtabular}
\end{table*}

Of two possible polar phases resulting from the condensation of the unstable
ferroelectric $E$~mode, the $Cm$~phase with atomic displacements along
the [110] direction has the lowest energy in all systems except for
KNbO$_3$/LaAlO$_3$, PbTiO$_3$/LaGaO$_3$, and SrTiO$_3$/BiScO$_3$. The
predominant [111] displacements correlate
with the fact that at normal conditions, bulk BaTiO$_3$ and KNbO$_3$ are
polarized along the [111] direction. The same polarization direction is also
characteristic of BiScO$_3$ when the octahedral rotations are turned off.
As an example, the energies of different low-symmetry phases for the
KNbO$_3$/BaTiO$_3$ superlattice are given in Table~\ref{table1}. It is seen that
the structures obtained from the condensation of $Z_5$, $R_4$, and $X_4$~unstable
phonons have a higher energy as compared to the energy of the $Cm$~phase.
The lattice parameters and atomic positions in ground-state structures of this
and other superlattices studied in this work are given in Tables~S2--S26
of the Supplementary data.

The absence of the ferroelectric instability in the $P4mm$~phase of
PbTiO$_3$/KTaO$_3$ and PbTiO$_3$/LaAlO$_3$ SLs can be explained by a tendency
of bulk PbTiO$_3$ to be polarized along the [001] axis. This instability
was also absent in SrTiO$_3$/LaAlO$_3$ and KTaO$_3$/LaAlO$_3$ superlattices,
in which both constituents are nonpolar. In addition, the ferroelectric
instability did not appear in the BaTiO$_3$/LaAlO$_3$ superlattice, in which
its absence is a result of the strong (by 2.3\%) in-plane compression of
the BaTiO$_3$ layers. Our calculations of the influence of strain on the
ground-state structure of BaTiO$_3$ showed that the biaxial compression
of 1\% is already sufficient for the $P4mm$~phase to become the most stable
polar phase in this material, in agreement with~\cite{PhysRevLett.80.1988,
PhysRevB.69.212101}.

Along with the ferroelectric instability, in the high-symmetry $P4mm$~phase
of twelve superlattices with the polar discontinuity (except for
KNbO$_3$/PbTiO$_3$, KNbO$_3$/BaTiO$_3$, KNbO$_3$/BaZrO$_3$, BaTiO$_3$/KTaO$_3$,
and KTaO$_3$/PbTiO$_3$), an antiferrodistortive instability with octahedra
rotations around the $z$~axis is observed. This is clearly seen by the
appearance of unstable phonons on the $M$--$A$ line (Fig.~\ref{fig3}).
The appearance of the AFD instability in a superlattice clearly correlates
with its existence in one or both components of the SL (SrTiO$_3$, SrZrO$_3$,
LaAlO$_3$, BiScO$_3$). A comparison of the energies of phases resulting from
the condensation of phonons at $M$ and $A$~points of the Brillouin zone shows
that of two phases resulting from the $M_4$~phonon condensation (space
group $P4bm$) and from the $A_4$~phonon condensation (space group $I4cm$),
the $P4bm$~phase was energetically more favorable in all cases except for
the PbTiO$_3$/BiScO$_3$ superlattice.

The most complex picture is observed in superlattices, in which both the
ferroelectric and AFD instabilities are simultaneously present in the
$P4mm$~phase. Since the energy difference between the $P4bm$ and $I4cm$~phases
is small and both phases can exhibit the ferroelectric instability, it was
necessary to consider all polar subgroups of these phases when searching
for the ground state. The KNbO$_3$/SrZrO$_3$ superlattice is an example of
such a system. Its phonon spectrum is shown in Fig.~\ref{fig4}, and the
energies of different phases are given in Table~\ref{table1}.

The calculations show that the ferroelectric instability of the $P4bm$ and
$I4cm$~phases is characteristic of KNbO$_3$/SrTiO$_3$, KNbO$_3$/SrZrO$_3$,
SrTiO$_3$/KTaO$_3$ superlattices and all SLs containing BiScO$_3$. Of two phases,
$Cm(2)$ and $Pc$, into which the $P4bm$ structure can transform upon the polar
distortion, the $Pc$~phase with polarization along the [110] direction of the
pseudocubic cell had lower energy in all SLs except for PbTiO$_3$/LaGaO$_3$.
Of two phases, $Cc$ and one more phase of the $Cm$~symmetry, $Cm(3)$, into which
the $I4cm$ structure can transform upon the polar distortion, the $Cc$~phase whose
polarization is also directed along the [110] direction of the pseudocubic cell
usually had a lower energy. In BiScO$_3$/LaAlO$_3$ and KNbO$_3$/BiScO$_3$
superlattices, it was the $Cm(3)$~phase which had a lower energy, but this
phase has never become the ground state. In KNbO$_3$/SrTiO$_3$ and
SrTiO$_3$/KTaO$_3$ superlattices, the energy difference between the $Cc$ and
$Pc$~phases was only 0.5--0.7~meV, whereas in other systems it was within
8--80~meV. Since the calculations proved the stability of both $Cc$ and
$Pc$~phases
in the two superlattices, the $Cc$~phase should be considered as a metastable
one. As the energy difference between the two phases is very small, there is
a real possibility that both structures can occur in an experiment simultaneously.
That is why the properties of these superlattices were calculated below for both
phases, $Pc$ and $Cc$.

The PbTiO$_3$/LaGaO$_3$ superlattice is the only one in which the $Cm(2)$~phase
is the ground state. It originates from the $P4bm$~phase and has the polarization
directed along the $[x0z]$ direction of the pseudocubic cell.

In addition to the unstable $M_4$~mode discussed above, in phonon spectra of
the $P4mm$~phase of nine SLs (KNbO$_3$/SrTiO$_3$, KNbO$_3$/SrZrO$_3$,
SrTiO$_3$/LaAlO$_3$, PbTiO$_3$/LaAlO$_3$, PbTiO$_3$/LaGaO$_3$, and all SLs
containing BiScO$_3$) one more doubly degenerate unstable $M_5$~mode is observed
(Fig.~\ref{fig4}). This mode describes the octahedra rotations around one or
both of the $x$ and $y$~axes. The distortions described by this mode with
the order parameters of ($\eta$, 0) and ($\eta$, $\eta$) result in the $Pma2$
and $Cmm2$ phases whose energies are usually higher than that of the $P4bm$~phase
(except for PbTiO$_3$/LaGaO$_3$, SrTiO$_3$/BiScO$_3$,
KNbO$_3$/BiScO$_3$, and KNbO$_3$/SrZrO$_3$ SLs). Both the $Pma2$ and $Cmm2$
phases are characterized by the ferroelectric instability. An analysis of the
polar subgroups of these phases (with $Pc$, $Cm$, and $Pm$ space groups) shows
that the $Pc$~phase has the lowest energy among them. It should be noted,
however, that in the BiScO$_3$/LaAlO$_3$ and PbTiO$_3$/LaGaO$_3$ superlattices,
the structure of this phase differs from that of the $Pc$~phase, which
originates from the $P4bm$~phase. In the BiScO$_3$/LaAlO$_3$ superlattice,
this new $Pc(2)$ phase becomes the ground-state structure because its energy is
lower than that of the $Pc$~phase originating from the $P4bm$~phase. In the
PbTiO$_3$/LaGaO$_3$ superlattice, the energy of the $Pc(2)$~phase is intermediate
between those of the $Pc$~phase and ground-state $Cm(2)$~structure. As the
$Pc(2)$~phase demonstrates its stability with respect to all acoustic and
optical distortions, we conclude that this is a metastable phase.

In two latter superlattices, BiScO$_3$/LaAlO$_3$ and PbTiO$_3$/LaGaO$_3$,
we encounter a situation when the metastable solutions appear. The cause of
this metastability is that the relaxation paths of the $P4bm$ and $Pma2$
($Cmm2$) phases, in which the rotations are described by the $M_4$ and $M_5$
irreducible representations, are separated by a potential barrier. In other
superlattices, these barriers are absent, and the structures, regardless
of the initial rotation pattern, relax to the same ground-state $Pc$~structure.
It may seem strange why the same phase appears as a result of condensation of
distortions described by two different irreducible representations. However,
it should be taken into account that the vertical axis of the octahedra in
the $Pc$~phase is slightly inclined, which means that, in fact, this phase is
described by two non-zero rotations around the coordinate axes. This explains
why two structures, in which rotational distortions are described by different
irreducible representations, relax to the same phase when the polar distortions
are turned on.

In SrTiO$_3$/LaAlO$_3$ and PbTiO$_3$/LaAlO$_3$ superlattices, both the $Pma2$
and $Cmm2$ phases relax to the $P4bm$~structure when the polar distortions
are turned on. Here, as in the case of the KNbO$_3$/LaAlO$_3$ SL, the strong
AFD instability of LaAlO$_3$ results in a complete suppression of the
ferroelectric instability which was present in the high-symmetry $P4mm$~phase.

In all BiScO$_3$-containing superlattices with the polar discontinuity, one
more unstable mode, $A_5$, appears in the phonon spectrum at the $A$~point.
An analysis of chains of phases induced by the corresponding distortions
showed that all obtained phases have the energy much higher than that of the
ground-state structure.

\begin{table}
\caption{\label{table2}Calculated polarization and energy gain resulting from
the in-plane ferroelectric distortion for all studied short-period superlattices.}
\begin{ruledtabular}
\begin{tabular}{cccccc}
Superlattice        & Space   & $P_x$ & $P_y$ & $P_z$    & $\Delta E$ \\
                    & group   & \multicolumn{3}{c}{(C/m$^2$)} & (meV) \\
\hline
KNbO$_3$/PbTiO$_3$  & $Cm$    & 0.181 & 0     &    0.567 &  3.3 \\
KNbO$_3$/BaTiO$_3$  & $Cm$    & 0.338 & 0     & $-$0.033 & 38.6 \\
KNbO$_3$/BaZrO$_3$  & $Cm$    & 0.229 & 0     &    0.181 & 33.8 \\
KNbO$_3$/SrTiO$_3$  & $Pc$    & 0.281 & 0     &    0.066 & 53.3;~9.0\footnotemark[1] \\
KNbO$_3$/SrZrO$_3$  & $Pc$    & 0.268 & 0     &    0.187 & 51.9;~49.8\footnotemark[1] \\
\hline
BaTiO$_3$/KTaO$_3$  & $Cm$    & 0.180 & 0     &    0.130 & 9.7 \\
BaTiO$_3$/LaAlO$_3$ & $P4bm$  & 0     & 0     &    0.057 & --- \\
\hline
SrTiO$_3$/KTaO$_3$  & $Pc$    & 0.105 & 0     &    0.141 & 11.9;~0.8\footnotemark[1] \\
SrTiO$_3$/LaAlO$_3$ & $P4bm$  & 0     & 0     &    0.012 & --- \\
SrTiO$_3$/BiScO$_3$ & $Pc$    & 0.501 & 0     &    0.329 & 671.1;~331.1\footnotemark[1] \\
\hline
PbTiO$_3$/KTaO$_3$  & $P4mm$  & 0     & 0     &    0.454 & --- \\
PbTiO$_3$/LaAlO$_3$ & $P4bm$  & 0     & 0     &    0.119 & --- \\
PbTiO$_3$/LaGaO$_3$ & $Cm(2)$ & 0.073 & 0     & $-$0.192 & 24.0\footnotemark[1] \\
PbTiO$_3$/BiScO$_3$ & $Pc$    & 0.710 & 0     &    0.330 & 865.5;~349.3\footnotemark[1] \\
\hline
KNbO$_3$/LaAlO$_3$  & $P4bm$  & 0     & 0     &    0.061 & --- \\
KTaO$_3$/LaAlO$_3$  & $P4bm$  & 0     & 0     &    0.109 & --- \\
KNbO$_3$/BiScO$_3$  & $Pc$    & 0.558 & 0     &    0.241 & 381.3;~281.5\footnotemark[1] \\
\hline
KNbO$_3$/NaTaO$_3$  & $Pc$    & 0.302 & 0     &    0.119 & 26.5;~26.4\footnotemark[1] \\
BaTiO$_3$/SrZrO$_3$ & $Pc$    & 0.234 & 0     &    0.072 & 186.8 \\
BaTiO$_3$/SrSnO$_3$ & $Pc$    & 0.189 & 0     &    0.072 & 275.9 \\
BiScO$_3$/LaAlO$_3$ & $Pc(2)$ & 0.493 & 0     & $-$0.116 & 535.1;~189.3\footnotemark[1] \\
\hline
BaTiO$_3$/SrTiO$_3$ & $Cm$    & 0.233 & 0     &    0.061 &  4.8 \\
KNbO$_3$/KTaO$_3$   & $Cm$    & 0.223 & 0     &    0.119 & 11.7 \\
PbTiO$_3$/PbZrO$_3$ & $Pc$    & 0.586 & 0     &    0.314 & 488.2;~147.4\footnotemark[1] \\
KNbO$_3$/NaNbO$_3$  & $Pmc2_1$ & 0.502 & 0    &    0     & 104.5;~98.6\footnotemark[2] \\
\end{tabular}
\end{ruledtabular}
\footnotetext[1]{As compared to the non-polar $P4bm$~phase.}
\footnotetext[2]{As compared to the non-polar $P4/mbm$~phase.}
\end{table}

The space groups of the energetically most favorable phases obtained for all
studied superlattices are given in Table~\ref{table2}. It is seen that in
superlattices exhibiting only the ferroelectric instability in the high-symmetry
$P4mm$~phase, the $Cm$~phase is the ground state. In superlattices exhibiting
only the AFD instability, the $P4bm$~phase is the ground state. And, finally,
in superlattices in which both instabilities are present, the $Pc$~phase is
the ground state, with the exception of the KNbO$_3$/LaAlO$_3$ and PbTiO$_3$/LaGaO$_3$
superlattices. In the first of these exceptions, the AFD instability suppresses
the ferroelectric one, and the $P4bm$~phase becomes the ground state. In the second
of them, the ground-state structure is $Cm(2)$. The discrepancy between our
results and those obtained for the KNbO$_3$/BaTiO$_3$ SL in \cite{PhysRevB.87.085305}
stems from the fact that the calculations in Ref.~\cite{PhysRevB.87.085305} were
performed for the superlattice clamped on the SrTiO$_3$ substrate; at these
boundary conditions, the $P4mm$~phase is indeed the ground state.

Table~\ref{table2} also presents the energy gain $\Delta E$ per 10-atom formula
unit, which results from the in-plane ferroelectric distortion. The obtained
values show that, at room temperature, the predicted ground-state structures
are likely to be observed in the KNbO$_3$/BaTiO$_3$, KNbO$_3$/BaZrO$_3$,
KNbO$_3$/SrZrO$_3$, SrTiO$_3$/BiScO$_3$, PbTiO$_3$/BiScO$_3$, and
KNbO$_3$/BiScO$_3$ superlattices. For the ground-state structures of all
superlattices, the calculated spontaneous polarization and components of the
piezoelectric tensor are presented in the following sections.

\subsubsection{Calculation of polarization}

Classical electrostatics of an electrically neutral interface between two
dielectrics requires that the electric displacement field components normal to
this interface are equal in two materials. If the materials are ferroelectrics
and have different spontaneous polarizations, a bound electric charge appears
at the interface, and the electric field generated by it makes the electric
displacement fields equal in two materials.

In superlattices with the polar discontinuity, a violation of the order of
charged $A$O and $B$O$_2$ planes generates an additional electrostatic perturbation
at the interfaces and creates a polarization jump of $\Delta P = e / 2a^2$ in
superlattices with a change in ionic charges of components by~$e$ and
$\Delta P = e / a^2$ in superlattices with a change in the ionic charges by
$2e$ (here $a$ is the in-plane lattice parameter of the
superlattice)~\cite{PhysRevB.79.100102}. An elegant solution to this problem
within the modern theory of polarization, which is valid in the general case,
was proposed in~\cite{PhysRevB.80.241103}. An application of this approach to
our superlattices enables us to calculate the electric displacement field in
them and to use it to determine the \emph{average} polarization in the SLs. In
this work, we are interested precisely in this property. The polarization values
in individual layers of a SL can be calculated by correcting the obtained average
polarization taking into account the jump in the ionic contribution to the Berry
phase at the interface and dielectric constants of individual components.%
    \footnote{The periodicity of a superlattice assumes that the electric field
    strengths ${\cal E}_1$ and ${\cal E}_2$ in its layers satisfy the condition
    ${\cal E}_1 x_1 + {\cal E}_2 x_2 = 0$, where $x_1$ and $x_2$ are the
    thicknesses of individual layers. This condition, combined with the equation
    $(P_1 + \epsilon_1 {\cal E}_1) - (P_2 + \epsilon_2 {\cal E}_2) = \Delta P$,
    where $P_1$, $P_2$, $\epsilon_1$, and $\epsilon_2$ are spontaneous
    polarizations and dielectric constants of two layers, gives a solution to
    this problem in the linear approximation. Solving a nonlinear problem
    requires more complex calculations.}

When calculating the polarization by the Berry phase method, it should be borne
in mind that the ionic contributions to the Berry phase are different in
nonpolar $Pm{\bar 3}m$ phases of I-V, II-IV, and III-III
perovskites~\cite{PhysicsofFerroelectrics2}. That is why for each SL with the
polar discontinuity it is necessary first to find the Berry phase for the nonpolar
structure before calculating the polarization. Unfortunately, in our case,
the calculation of the $z$~component of the Berry phase becomes a problem
because it is impossible to reverse the polarization or construct a non-polar
structure for the high-symmetry $P4mm$~phase, which does not have $\sigma_z$
mirror plane. To \emph{estimate} $P_z$, we considered unrelaxed structures
with ideal atomic positions corresponding to the cubic perovskite structure in
both layers. In this structure, the electron contribution to the Berry phase
is nonzero because of the redistribution of the electron density between the
layers, and the ionic contribution reflects the difference in the Berry
phases of individual perovskites. The average polarization in a superlattice
was calculated using standard formulas from a change in the Berry phase upon
the transition from the above-described unrelaxed structure to the ground-state one.
To correctly determine the polarization from the change in the Berry phase,
which is well-defined modulo $2\pi$, for all superlattices the calculations
were also performed for at least one intermediate point at which the atoms
are halfway between the ground-state and unrelaxed structures.

The calculated polarizations are given in Table~\ref{table2}. The components
of the polarization vector are given relative to the axes of the standard
crystallographic settings for tetragonal and monoclinic unit cells (their axes
are often rotated in the $xy$~plane relative to each other by 45$^\circ$).
A comparison of polarizations calculated for structures with and without
octahedral rotations shows that in SLs with the $Pc$ ground-state structure,
the neglect of these rotations can result in an error in determining $P_z$
up to 30\% and, in some cases, even in an error in the sign of this quantity.
For the $P4mm$~phase, our values reasonably agree with the published values
of $P_z = 0.532$~C/m$^2$ for PbTiO$_3$/KNbO$_3$~\cite{ChinPhysB.27.027701},
$P_z = 0.202$~C/m$^2$ for PbTiO$_3$/LaAlO$_3$~\cite{ChinPhysB.27.027701}, and
$P_z = 0.38$~C/m$^2$ for PbTiO$_3$/KTaO$_3$~\cite{ChinPhysLett.33.026302} SLs.

\subsubsection{Piezoelectric properties}

The reason for our interest to ferroelectric instability in superlattices
with the polar discontinuity is that in such SLs it is possible to obtain rather
high piezoelectric coefficients resulting from the in-plane ferroelectric
phase transitions appearing in them. The literature data on the piezoelectric
properties of such superlattices are limited to calculations for the
high-symmetry $P4mm$~phase
of PbTiO$_3$/LaAlO$_3$ and KNbO$_3$/PbTiO$_3$ SLs~\cite{JApplPhys.109.066107,
ChinPhysB.27.027701} and of the PbTiO$_3$/KTaO$_3$ one~\cite{ChinPhysLett.33.026302}.
As the ground-state structure of the first two superlattices differs from $P4mm$,
more correct calculations for these SLs are needed. For other superlattices
considered in this work, data on their piezoelectric properties are absent.

\begin{table*}
\caption{\label{table3}Nonzero components of the piezoelectric tensor $d_{i\nu}$
(in~pC/N) for the ground-state structures of all studied short-period superlattices.}
\begin{ruledtabular}
\begin{tabular}{cccccccccccc}
Superlattice         & $d_{11}$ & $d_{12}$ & $d_{13}$ & $d_{15}$ & $d_{24}$ & $d_{26}$ & $d_{31}$ & $d_{32}$ & $d_{33}$ & $d_{35}$ \\
\hline
KNbO$_3$/PbTiO$_3$   &  13.4  &    3.7   & $-$13.9  &   158.3  &   273.3  &  137.3   &  $-$3.0  &  $-$2.9  & $-$10.9  & $-$0.8 \\
KNbO$_3$/BaTiO$_3$   &  23.5  &    9.6   & $-$12.7  &     3.8  &  $-$2.8  &  106.4   &    10.7  &     7.9  & $-$16.3  &   17.4 \\
KNbO$_3$/BaZrO$_3$   &  16.3  &    4.7   & $-$11.7  &     5.6  &     7.9  &   20.8   &  $-$8.9  &  $-$7.3  &    16.7  &    1.8 \\
KNbO$_3$/SrTiO$_3$   &  35.9  &   15.3   & $-$19.1  &     8.3  &    41.7  &  151.0   & $-$22.0  & $-$12.1  &    16.1  &    9.4 \\
KNbO$_3$/SrZrO$_3$   &  30.8  &   10.1   & $-$18.8  &     9.1  &    11.0  &   85.5   & $-$18.1  &  $-$9.8  &    17.7  &   12.9 \\
\hline
BaTiO$_3$/KTaO$_3$   &  19.2  &    9.5   & $-$18.5  &     2.5  &     7.9  &   25.3   & $-$18.1  & $-$11.6  &    43.4  &   10.1 \\
BaTiO$_3$/LaAlO$_3$  &   ---  &    ---   &     ---  &     1.4  &     1.4  &    ---   &     3.0  &     3.0  &  $-$4.6  &    --- \\
\hline
SrTiO$_3$/KTaO$_3$   &  39.7  &   22.8   & $-$27.6  &    28.2  &    46.9  &   41.2   & $-$18.4  & $-$13.0  &    26.1  & $-$4.2 \\
SrTiO$_3$/LaAlO$_3$  &   ---  &    ---   &     ---  &     2.0  &     2.0  &    ---   &     1.0  &     1.0  &  $-$1.6  &    --- \\
SrTiO$_3$/BiScO$_3$  & 23.6   &    5.4   &  $-$6.0  &    14.3  &    17.9  &   24.6   &  $-$4.7  &     0.6  &    11.8  &   12.2 \\
\hline
PbTiO$_3$/KTaO$_3$   &   ---  &    ---   &     ---  &    91.4  &    91.4  &    ---   &  $-$8.8  &  $-$8.8  &    41.2  &    --- \\
PbTiO$_3$/LaAlO$_3$  &   ---  &    ---   &     ---  &   158.6  &   158.6  &    ---   &  $-$4.3  &  $-$4.3  &  $-$2.8  &    --- \\
PbTiO$_3$/LaGaO$_3$  &  19.1  & $-$5.8   &  $-$3.0  & $-$29.6  &    15.1  & $-$35.1  &     7.8  &  $-$2.9  & $-$18.5  &    4.3 \\
PbTiO$_3$/BiScO$_3$  &  31.1  &    2.7   &  $-$7.4  &     5.1  &     1.5  &   47.0   &  $-$2.3  &  $-$4.0  &    18.5  &   13.6 \\
\hline
KNbO$_3$/LaAlO$_3$   &   ---  &    ---   &     ---  &     5.7  &     5.7  &    ---   &     1.1  &     1.1  &  $-$0.8  &    --- \\
KTaO$_3$/LaAlO$_3$   &   ---  &    ---   &     ---  &     5.3  &     5.3  &    ---   &  $-$0.8  &  $-$0.8  &     3.7  &    --- \\
KNbO$_3$/BiScO$_3$   &  19.2  &    9.2   &  $-$5.9  &    10.0  &    23.2  &   51.2   &  $-$0.5  &     3.9  &     2.1  &    6.9 \\
\hline
KNbO$_3$/NaTaO$_3$   &  90.2  &   19.8   & $-$97.1  &    24.3  & $-$16.0  &   70.2   &   154.6  &    43.1  & $-$209.2 &   73.5 \\
BaTiO$_3$/SrZrO$_3$  &  23.3  &    4.9   & $-$12.8  &     1.2  &     2.6  &   19.3   &     3.9  &     4.0  &  $-$7.8  &    5.1 \\
BaTiO$_3$/SrSnO$_3$  &  23.9  &    6.6   & $-$12.7  &  $-$7.1  &  $-$6.4  &   23.2   &  $-$1.0  &  $-$2.2  &     3.2  &    2.1 \\
BiScO$_3$/LaAlO$_3$  &  30.2  & $-$1.1   &  $-$5.9  &  $-$5.7  &     8.3  &   23.1   &     0.3  &     1.3  &  $-$3.5  &   10.5 \\ 
\hline
BaTiO$_3$/SrTiO$_3$  & 103.5  &   37.4   & $-$117.1 & $-$59.5  &    21.1  &  187.9   & $-$287.3 & $-$131.5 &   460.5  &  288.1 \\
KNbO$_3$/KTaO$_3$    & 108.3  &   60.4   & $-$157.1 & $-$15.9  &    18.7  &  123.9   & $-$181.6 & $-$111.3 &   315.5  &   45.4 \\
PbTiO$_3$/PbZrO$_3$  &  62.1  &   12.5   & $-$38.8  &     3.2  &    29.9  &  132.7   & $-$45.8  & $-$19.5  &    70.4  &   26.6 \\
KNbO$_3$/NaNbO$_3$   &  17.9  &    5.3   &  $-$9.2  &     ---  &     ---  &  153.0   &    ---   &    ---   &    ---   &   60.1 \\
\end{tabular}
\end{ruledtabular}
\end{table*}

In superlattices with a tetragonal ground-state structure, in which the
polarization is directed along the $z$~axis, five components of the piezoelectric
tensor $d_{i\nu}$ are nonzero. Among them, the highest values of $d_{i\nu}$ are
obtained for PbTiO$_3$/KTaO$_3$ and PbTiO$_3$/LaAlO$_3$ SLs (Table~\ref{table3}).
Interestingly, among these coefficients, the $d_{15}$ values turned out to be
the largest. This coefficient characterizes the polarization $P_x$ that appears
as a result of the $xz$ shear strain of the unit cell, that is, as a result of
the inclination of the polarization vector. However, no clear correlation between
$d_{15}$ and $P_z$ was observed in tetragonal SLs.
Moreover, in the related system, BaTiO$_3$/LaAlO$_3$,
the piezoelectric coefficients were unexpectedly low (Table~\ref{table3}). This
means that the inclination of the polarization vector is not an effective way
for obtaining high piezoelectric coefficients.

In superlattices with a monoclinic ground-state structure, in which the
polarization vector lies in the $xz$~plane, the piezoelectric tensor is characterized
by ten nonzero components. In these SLs, the highest values of $d_{i\nu}$
are the $d_{24}$ and $d_{26}$ coefficients, which describe the appearance of
polarization in the $y$~direction normal to the $xz$~plane under the $yz$ and
$xy$ shear strain. An analysis of the obtained data also does not find a clear
correlation between the piezoelectric coefficients and the average polarization.
Stretching the unit cells of these superlattices in the $xz$~plane changes the
polarization, but the corresponding piezoelectric coefficients ($d_{11}$ and
$d_{33}$ in Table~\ref{table3}) are not the largest.

\begin{table*}
\caption{\label{table4}The values of the $\partial u_{1}^i / \partial \sigma_{15}$
component of the $\partial u_{\alpha}^i / \partial \sigma_{\mu\nu}$ tensor (in~{\AA})
for all atoms in the ground-state structure of PbTiO$_3$/LaAlO$_3$ and BaTiO$_3$/LaAlO$_3$
superlattices ($P4bm$~phase) and in ordinary superlattices, PbTiO$_3$/PbZrO$_3$
($Pc$~phase) and KNbO$_3$/NaNbO$_3$ ($Pmc2_1$~phase), of systems known for their
piezoelectric properties. Atoms 1--10 are numbered according to Fig.~\ref{fig1};
atoms 11--20 are located in the second part of the doubled unit cell of the
high-temperature phase and are shifted from atoms 1--10 along the $x$~axis.
For atomic positions of all atoms see tables in the Supplementary data.}
\begin{ruledtabular}
\begin{tabular}{ccccc}
Atom~$i$ & PbTiO$_3$/LaAlO$_3$ & BaTiO$_3$/LaAlO$_3$ & PbTiO$_3$/PbZrO$_3$ & KNbO$_3$/NaNbO$_3$ \\
\hline
1   &   +5.18 &   +0.09 &   +0.35 &   +0.00 \\
2   &   +2.37 & $-$0.08 &   +0.03 & $-$0.58 \\
3   & $-$1.19 &   +0.24 &   +0.14 & $-$0.31 \\
4   & $-$5.61 &   +0.10 &   +0.17 & $-$0.51 \\
5   & $-$1.02 & $-$0.11 &   +0.49 &   +0.00 \\
6   &   +4.56 &   +0.73 &   +0.44 &   +0.00 \\
7   &   +0.76 &   +0.02 & $-$0.23 &   +0.58 \\
8   & $-$0.21 & $-$0.31 & $-$0.31 &   +0.31 \\
9   & $-$0.22 & $-$0.15 & $-$0.83 &   +0.51 \\
10  & $-$0.84 & $-$0.10 & $-$0.42 &   +0.00 \\
11  &   +1.71 & $-$0.38 &   +0.01 &   +0.00 \\
12  &   +2.37 & $-$0.08 &   +0.08 & $-$0.41 \\
13  & $-$1.19 &   +0.24 & $-$0.11 & $-$0.51 \\
14  & $-$5.61 &   +0.10 & $-$0.43 & $-$1.15 \\
15  & $-$1.02 & $-$0.11 &   +0.47 &   +0.00 \\
16  &   +0.48 &   +0.37 &   +0.76 &   +0.00 \\
17  &   +0.76 &   +0.02 & $-$0.19 &   +0.41 \\
18  & $-$0.21 & $-$0.31 & $-$0.06 &   +0.51 \\
19  & $-$0.22 & $-$0.15 & $-$0.26 &   +1.15 \\
20  & $-$0.84 & $-$0.10 & $-$0.12 &   +0.00 \\
\end{tabular}
\end{ruledtabular}
\end{table*}

To understand the mechanism of the appearance of high piezoelectric coefficients
in some superlattices with the polar discontinuity, we analyzed the third-rank
tensors $\partial u_{\alpha}^i / \partial \sigma_{\mu\nu}$. This tensor
characterizes the displacement of the $i$th atom in the unit cell in the
$\alpha$~direction produced by strain $\sigma_{\mu\nu}$. It turned out that
in SLs exhibiting strong piezoelectricity, the values of some components
of these tensors for some atoms reach 10--15~{\AA} (that is, the unit cell strain
of 1\% generates atomic displacements exceeding 0.1~{\AA}). For example, in the
PbTiO$_3$/LaAlO$_3$ superlattice, such atoms are Pb(1), La(6), and two oxygen atoms
O(4) and O(14) located in the TiO$_2$ layer (Table~\ref{table4}); the contributions
of O(8) and O(9) oxygen atoms located in the AlO$_2$ layer are 25~times smaller.

The fact that strong distortions in the PbTiO$_3$/LaAlO$_3$ SL are observed in
the TiO$_2$ layer suggests that the Ti(2) atom also actively participates
in the appearance of polarization. Indeed, the displacement pattern
of the Pb, Ti, and O atoms in the PbO and TiO$_2$ layers resembles that of the
polar mode in PbTiO$_3$. A more detailed analysis of atomic displacements in
monoclinically strained SL, whose deformation is described by a nonzero component
$\sigma_{xz} = 0.005$, simultaneously finds two types of distortions: the rotations
of corner-linked TiO$_6$ and AlO$_6$ octahedra around the $x$~axis, resembling
rotations described by an unstable phonon at the $R$~point of the cubic perovskite
structure, and out-of-phase
(polar) displacements of Pb(1), La(6), Ti(2), O(4), and O(14) atoms along the
$x$~axis, like in the polar mode in PbTiO$_3$. The observed displacement pattern
is very close (with the correlation coefficient of 0.902) to the eigenvector
of the soft $E$~mode with a surprisingly low frequency of 27~cm$^{-1}$ in the
phonon spectrum of the $P4bm$~phase of this SL. Under the influence of strain,
the irreducible representation~$E$ reduces to a sum of $A' + A''$ irreducible
representations of the low-symmetry $Cm$~phase, and the structure distorts
according to the eigenvector of the full-symmetry $A'$~mode. In other SLs with
lower piezoelectric properties, the frequencies of the corresponding modes were
higher (40--120~cm$^{-1}$), but we didn't find simple correlations between
the frequencies and effective charges of these modes and the piezoelectric
properties. This question needs more detailed investigation.

\begin{table*}
\caption{\label{table5}Nonzero components of the piezoelectric tensor $d_{i\nu}$
(in~pC/N) in the ground and metastable states of two short-period superlattices
with the polar discontinuity.}
\begin{ruledtabular}
\begin{tabular}{cccccccccccc}
Phase & $d_{11}$ & $d_{12}$ & $d_{13}$ & $d_{15}$ & $d_{24}$ & $d_{26}$ & $d_{31}$ & $d_{32}$ & $d_{33}$ & $d_{35}$ \\
\hline
\multicolumn{11}{c}{KNbO$_3$/SrTiO$_3$ superlattice} \\
$Pc$  &  35.9    & 15.3     & $-$19.1  &   8.3    &   41.7   &  151.0   & $-$22.0  & $-$12.1  &    16.1  &    9.4 \\
$Cc$  &  35.0    & 14.8     & $-$18.3  &   7.7    &   41.4   &  154.0   & $-$21.4  & $-$11.6  &    15.7  &   11.1 \\
\hline
\multicolumn{11}{c}{SrTiO$_3$/KTaO$_3$ superlattice} \\
$Pc$  &  39.7    & 22.8     & $-$27.6  &  28.2    &   46.9   &   41.2   & $-$18.4  & $-$13.0  &    26.1  & $-$4.2 \\
$Cc$  &  39.2    & 22.3     & $-$27.4  &  27.1    &   45.7   &   41.1   & $-$18.6  & $-$13.1  &    26.3  & $-$3.7 \\
\end{tabular}
\end{ruledtabular}
\end{table*}

A comparison of the piezoelectric properties of the metastable $Cc$~phase and
the ground-state $Pc$~phase in KNbO$_3$/SrTiO$_3$ and SrTiO$_3$/KTaO$_3$
superlattices, in which the energy difference between the two structures
is very small, shows that their piezoelectric tensors are quite close to each
other (Table~\ref{table5}). This means that a change in the character of the
octahedra rotations weakly affects the piezoelectric properties of SLs.

According to our calculations, the piezoelectric coefficients in studied
superlattices can reach 150--270~pC/N. A comparison of the obtained results with
the published data in some cases finds their good agreement. For example, for
the $P4mm$~phase of PbTiO$_3$/LaAlO$_3$ SL our result $e_{33} = -3.52$~C/m$^2$
is quite close to the value $e_ {33} = -2.85$~C/m$^2$ obtained in \cite{JApplPhys.109.066107}.
However, our value of $d_{33} = -18.9$~pC/N for the same phase of the same SL
differs from the value $d_{33} = 13.9$~pC/N obtained in \cite{ChinPhysB.27.027701}:
these values are close in magnitude, but have a different sign. An additional
reason for the stronger discrepancies here may be a neglect of the difference in
the displacements and effective charges of the O atoms in Ref.~\cite{ChinPhysB.27.027701}
(according to our data, the effective charges varies from $-$2.07 to $-$5.84).
As for the PbTiO$_3$/KTaO$_3$ SL studied in \cite{ChinPhysLett.33.026302}, the
$e_{33}$ and $e_{15}$ piezoelectric coefficients obtained there are almost
100~times less than our data, and the $e_{31}$~value is close to our result in
magnitude, but differs in sign.

In conclusion, it was interesting to compare the piezoelectric properties of
superlattices with the polar discontinuity with those of ``isoelectronic''
superlattices: (1)~SLs without the polar discontinuity, in which the broken
symmetry and high-symmetry $P4mm$~structure are formed, by analogy with
\cite{PhysRevLett.84.5636}, as a result of the absence of the mirror symmetry
in the sequence of layers (BaTiO$_3$/SrZrO$_3$, BaTiO$_3$/SrSnO$_3$,
KNbO$_3$/NaTaO$_3 $, and BiScO$_3$/LaAlO$_3$), and (2)~ordinary SLs with
one substituted
atom (BaTiO$_3$/SrTiO$_3$, KNbO$_3$/KTaO$_3$, and two famous systems known for
their high piezoelectric properties, PbTiO$_3$/PbZrO$_3$ and ``lead-free''
KNbO$_3$/NaNbO$_3$). The piezoelectric properties of these superlattices are
given in the last eight lines of Table~\ref{table3}. It is seen that in
spite of a large scatter of the data, superlattices with the polar discontinuity
are not much different in their piezoelectric properties from other
superlattices (high piezoelectric properties of
BaTiO$_3$/SrTiO$_3$ and KNbO$_3$/KTaO$_3$ SLs, as follows from the corresponding
values of $\Delta E$ (Table~\ref{table2}), can be realized only at low
temperatures). Thus, the only advantage of superlattices with the polar
discontinuity as well as of SLs with broken symmetry is that one component of
polarization in them is not switchable.

\section{Conclusions}

First-principles calculations have been used to study the stability of a
high-symmetry $P4mm$~polar phase in seventeen ferroelectric perovskite
superlattices with the polar discontinuity. In most superlattices, this phase
exhibits either the ferroelectric instability or the antiferrodistortive one,
or both of them simultaneously. For each superlattice, the ground-state
structure, the structure of possible metastable phases, spontaneous polarization,
and piezoelectric properties were calculated. A~comparison of the piezoelectric
properties of superlattices with the polar discontinuity and those without
the polar discontinuity (superlattices with broken symmetry and ordinary
superlattices) showed that these properties are not much different. It was
demonstrated that high piezoelectric coefficients (up to 150--270~pC/N) in some
superlattices with the polar discontinuity are due to the appearance of strong
lattice distortions, whose symmetry follows that of a low-lying polar phonon
mode of the ground-state structure under the influence of external strain.

\bigskip

\textbf{Data availability statement}

\bigskip
All data included in this study are available upon request from the author.

\bigskip
\textbf{Declaration of competing interests}

\bigskip
The author declares that he has no known competing financial interests or
personal relationships that could have appeared to influence the work reported
in this paper.

\mbox{ }

\begin{acknowledgments}
This work was partially supported by the Russian Foundation for Basic Research
under Grant 17-02-01068.
\end{acknowledgments}

\appendix

\section{Supplementary data}

Supplementary data associated with this article can be found in the online
version at \texttt{https://doi.org/10.1016/j.commatsci.2020.110113}.


\begin{thebibliography}{86}%
\makeatletter
\providecommand \@ifxundefined [1]{%
 \@ifx{#1\undefined}
}%
\providecommand \@ifnum [1]{%
 \ifnum #1\expandafter \@firstoftwo
 \else \expandafter \@secondoftwo
 \fi
}%
\providecommand \@ifx [1]{%
 \ifx #1\expandafter \@firstoftwo
 \else \expandafter \@secondoftwo
 \fi
}%
\providecommand \natexlab [1]{#1}%
\providecommand \enquote  [1]{``#1''}%
\providecommand \bibnamefont  [1]{#1}%
\providecommand \bibfnamefont [1]{#1}%
\providecommand \citenamefont [1]{#1}%
\providecommand \href@noop [0]{\@secondoftwo}%
\providecommand \href [0]{\begingroup \@sanitize@url \@href}%
\providecommand \@href[1]{\@@startlink{#1}\@@href}%
\providecommand \@@href[1]{\endgroup#1\@@endlink}%
\providecommand \@sanitize@url [0]{\catcode `\\12\catcode `\$12\catcode
  `\&12\catcode `\#12\catcode `\^12\catcode `\_12\catcode `\%12\relax}%
\providecommand \@@startlink[1]{}%
\providecommand \@@endlink[0]{}%
\providecommand \url  [0]{\begingroup\@sanitize@url \@url }%
\providecommand \@url [1]{\endgroup\@href {#1}{\urlprefix }}%
\providecommand \urlprefix  [0]{URL }%
\providecommand \Eprint [0]{\href }%
\providecommand \doibase [0]{https://doi.org/}%
\providecommand \selectlanguage [0]{\@gobble}%
\providecommand \bibinfo  [0]{\@secondoftwo}%
\providecommand \bibfield  [0]{\@secondoftwo}%
\providecommand \translation [1]{[#1]}%
\providecommand \BibitemOpen [0]{}%
\providecommand \bibitemStop [0]{}%
\providecommand \bibitemNoStop [0]{.\EOS\space}%
\providecommand \EOS [0]{\spacefactor3000\relax}%
\providecommand \BibitemShut  [1]{\csname bibitem#1\endcsname}%
\let\auto@bib@innerbib\@empty
\bibitem [{\citenamefont {Iijima}\ \emph {et~al.}(1992)\citenamefont {Iijima},
  \citenamefont {Terashima}, \citenamefont {Bando}, \citenamefont {Kamigaki},\
  and\ \citenamefont {Terauchi}}]{JApplPhys.72.2840}%
  \BibitemOpen
  \bibfield  {author} {\bibinfo {author} {\bibfnamefont {K.}~\bibnamefont
  {Iijima}}, \bibinfo {author} {\bibfnamefont {T.}~\bibnamefont {Terashima}},
  \bibinfo {author} {\bibfnamefont {Y.}~\bibnamefont {Bando}}, \bibinfo
  {author} {\bibfnamefont {K.}~\bibnamefont {Kamigaki}},\ and\ \bibinfo
  {author} {\bibfnamefont {H.}~\bibnamefont {Terauchi}},\ }\bibfield  {title}
  {\bibinfo {title} {Atomic layer growth of oxide thin films with
  perovskite-type structure by reactive evaporation},\ }\href
  {https://doi.org/10.1063/1.351536} {\bibfield  {journal} {\bibinfo  {journal}
  {J. Appl. Phys.}\ }\textbf {\bibinfo {volume} {72}},\ \bibinfo {pages} {2840}
  (\bibinfo {year} {1992})}\BibitemShut {NoStop}%
\bibitem [{\citenamefont {Tsurumi}\ \emph {et~al.}(1994)\citenamefont
  {Tsurumi}, \citenamefont {Suzuki}, \citenamefont {Yamane},\ and\
  \citenamefont {Daimon}}]{JapJApplPhys.33.5192}%
  \BibitemOpen
  \bibfield  {author} {\bibinfo {author} {\bibfnamefont {T.}~\bibnamefont
  {Tsurumi}}, \bibinfo {author} {\bibfnamefont {T.}~\bibnamefont {Suzuki}},
  \bibinfo {author} {\bibfnamefont {M.}~\bibnamefont {Yamane}},\ and\ \bibinfo
  {author} {\bibfnamefont {M.}~\bibnamefont {Daimon}},\ }\bibfield  {title}
  {\bibinfo {title} {Fabrication of barium titanate/strontium titanate
  artificial superlattice by atomic layer epitaxy},\ }\href
  {https://doi.org/10.1143/JJAP.33.5192} {\bibfield  {journal} {\bibinfo
  {journal} {Jap. J. Appl. Phys.}\ }\textbf {\bibinfo {volume} {33}},\ \bibinfo
  {pages} {5192} (\bibinfo {year} {1994})}\BibitemShut {NoStop}%
\bibitem [{\citenamefont {Tabata}\ \emph {et~al.}(1994)\citenamefont {Tabata},
  \citenamefont {Tanaka},\ and\ \citenamefont {Kawai}}]{ApplPhysLett.65.1970}%
  \BibitemOpen
  \bibfield  {author} {\bibinfo {author} {\bibfnamefont {H.}~\bibnamefont
  {Tabata}}, \bibinfo {author} {\bibfnamefont {H.}~\bibnamefont {Tanaka}},\
  and\ \bibinfo {author} {\bibfnamefont {T.}~\bibnamefont {Kawai}},\ }\bibfield
   {title} {\bibinfo {title} {Formation of artificial BaTiO$_3$/SrTiO$_3$
  superlattices using pulsed laser deposition and their dielectric
  properties},\ }\href {https://doi.org/10.1063/1.112837} {\bibfield  {journal}
  {\bibinfo  {journal} {Appl. Phys. Lett.}\ }\textbf {\bibinfo {volume} {65}},\
  \bibinfo {pages} {1970} (\bibinfo {year} {1994})}\BibitemShut {NoStop}%
\bibitem [{\citenamefont {Tabata}\ and\ \citenamefont
  {Kawai}(1997)}]{ApplPhysLett.70.321}%
  \BibitemOpen
  \bibfield  {author} {\bibinfo {author} {\bibfnamefont {H.}~\bibnamefont
  {Tabata}}\ and\ \bibinfo {author} {\bibfnamefont {T.}~\bibnamefont {Kawai}},\
  }\bibfield  {title} {\bibinfo {title} {Dielectric properties of strained
  (Sr,Ca)TiO$_3$/(Ba,Sr)TiO$_3$ artificial lattices},\ }\href
  {https://doi.org/10.1063/1.118202} {\bibfield  {journal} {\bibinfo  {journal}
  {Appl. Phys. Lett.}\ }\textbf {\bibinfo {volume} {70}},\ \bibinfo {pages}
  {321} (\bibinfo {year} {1997})}\BibitemShut {NoStop}%
\bibitem [{\citenamefont {Qu}\ \emph {et~al.}(1998)\citenamefont {Qu},
  \citenamefont {Evstigneev}, \citenamefont {Johnson},\ and\ \citenamefont
  {Prince}}]{ApplPhysLett.72.1394}%
  \BibitemOpen
  \bibfield  {author} {\bibinfo {author} {\bibfnamefont {B.~D.}\ \bibnamefont
  {Qu}}, \bibinfo {author} {\bibfnamefont {M.}~\bibnamefont {Evstigneev}},
  \bibinfo {author} {\bibfnamefont {D.~J.}\ \bibnamefont {Johnson}},\ and\
  \bibinfo {author} {\bibfnamefont {R.~H.}\ \bibnamefont {Prince}},\ }\bibfield
   {title} {\bibinfo {title} {Dielectric properties of BaTiO$_3$/SrTiO$_3$
  multilayered thin films prepared by pulsed laser deposition},\ }\href
  {https://doi.org/10.1063/1.121066} {\bibfield  {journal} {\bibinfo  {journal}
  {Appl. Phys. Lett.}\ }\textbf {\bibinfo {volume} {72}},\ \bibinfo {pages}
  {1394} (\bibinfo {year} {1998})}\BibitemShut {NoStop}%
\bibitem [{\citenamefont {Specht}\ \emph {et~al.}(1998)\citenamefont {Specht},
  \citenamefont {Christen}, \citenamefont {Norton},\ and\ \citenamefont
  {Boatner}}]{PhysRevLett.80.4317}%
  \BibitemOpen
  \bibfield  {author} {\bibinfo {author} {\bibfnamefont {E.~D.}\ \bibnamefont
  {Specht}}, \bibinfo {author} {\bibfnamefont {H.-M.}\ \bibnamefont
  {Christen}}, \bibinfo {author} {\bibfnamefont {D.~P.}\ \bibnamefont
  {Norton}},\ and\ \bibinfo {author} {\bibfnamefont {L.~A.}\ \bibnamefont
  {Boatner}},\ }\bibfield  {title} {\bibinfo {title} {X-ray diffraction
  measurement of the effect of layer thickness on the ferroelectric transition
  in epitaxial KTaO$_3$/KNbO$_3$ multilayers},\ }\href
  {https://doi.org/10.1103/PhysRevLett.80.4317} {\bibfield  {journal} {\bibinfo
   {journal} {Phys. Rev. Lett.}\ }\textbf {\bibinfo {volume} {80}},\ \bibinfo
  {pages} {4317} (\bibinfo {year} {1998})}\BibitemShut {NoStop}%
\bibitem [{\citenamefont {O'Neill}\ \emph {et~al.}(2000)\citenamefont
  {O'Neill}, \citenamefont {Bowman},\ and\ \citenamefont
  {Gregg}}]{ApplPhysLett.77.1520}%
  \BibitemOpen
  \bibfield  {author} {\bibinfo {author} {\bibfnamefont {D.}~\bibnamefont
  {O'Neill}}, \bibinfo {author} {\bibfnamefont {R.~M.}\ \bibnamefont
  {Bowman}},\ and\ \bibinfo {author} {\bibfnamefont {J.~M.}\ \bibnamefont
  {Gregg}},\ }\bibfield  {title} {\bibinfo {title} {Dielectric enhancement and
  Maxwell--Wagner effects in ferroelectric superlattice structures},\ }\href
  {https://doi.org/10.1063/1.1290691} {\bibfield  {journal} {\bibinfo
  {journal} {Appl. Phys. Lett.}\ }\textbf {\bibinfo {volume} {77}},\ \bibinfo
  {pages} {1520} (\bibinfo {year} {2000})}\BibitemShut {NoStop}%
\bibitem [{\citenamefont {Nakagawara}\ \emph {et~al.}(2000)\citenamefont
  {Nakagawara}, \citenamefont {Shimuta}, \citenamefont {Makino}, \citenamefont
  {Arai}, \citenamefont {Tabata},\ and\ \citenamefont
  {Kawai}}]{ApplPhysLett.77.3257}%
  \BibitemOpen
  \bibfield  {author} {\bibinfo {author} {\bibfnamefont {O.}~\bibnamefont
  {Nakagawara}}, \bibinfo {author} {\bibfnamefont {T.}~\bibnamefont {Shimuta}},
  \bibinfo {author} {\bibfnamefont {T.}~\bibnamefont {Makino}}, \bibinfo
  {author} {\bibfnamefont {S.}~\bibnamefont {Arai}}, \bibinfo {author}
  {\bibfnamefont {H.}~\bibnamefont {Tabata}},\ and\ \bibinfo {author}
  {\bibfnamefont {T.}~\bibnamefont {Kawai}},\ }\bibfield  {title} {\bibinfo
  {title} {Epitaxial growth and dielectric properties of (111) oriented
  BaTiO$_3$/SrTiO$_3$ superlattices by pulsed-laser deposition},\ }\href
  {https://doi.org/10.1063/1.1324985} {\bibfield  {journal} {\bibinfo
  {journal} {Appl. Phys. Lett.}\ }\textbf {\bibinfo {volume} {77}},\ \bibinfo
  {pages} {3257} (\bibinfo {year} {2000})}\BibitemShut {NoStop}%
\bibitem [{\citenamefont {Shimuta}\ \emph {et~al.}(2002)\citenamefont
  {Shimuta}, \citenamefont {Nakagawara}, \citenamefont {Makino}, \citenamefont
  {Arai}, \citenamefont {Tabata},\ and\ \citenamefont
  {Kawai}}]{JapplPhys.91.2290}%
  \BibitemOpen
  \bibfield  {author} {\bibinfo {author} {\bibfnamefont {T.}~\bibnamefont
  {Shimuta}}, \bibinfo {author} {\bibfnamefont {O.}~\bibnamefont {Nakagawara}},
  \bibinfo {author} {\bibfnamefont {T.}~\bibnamefont {Makino}}, \bibinfo
  {author} {\bibfnamefont {S.}~\bibnamefont {Arai}}, \bibinfo {author}
  {\bibfnamefont {H.}~\bibnamefont {Tabata}},\ and\ \bibinfo {author}
  {\bibfnamefont {T.}~\bibnamefont {Kawai}},\ }\bibfield  {title} {\bibinfo
  {title} {Enhancement of remanent polarization in epitaxial
  BaTiO$_3$/SrTiO$_3$ superlattices with ``asymmetric'' structure},\ }\href
  {https://doi.org/10.1063/1.1434547} {\bibfield  {journal} {\bibinfo
  {journal} {J. Appl. Phys.}\ }\textbf {\bibinfo {volume} {91}},\ \bibinfo
  {pages} {2290} (\bibinfo {year} {2002})}\BibitemShut {NoStop}%
\bibitem [{\citenamefont {Sigman}\ \emph {et~al.}(2002)\citenamefont {Sigman},
  \citenamefont {Norton}, \citenamefont {Christen}, \citenamefont {Fleming},\
  and\ \citenamefont {Boatner}}]{PhysRevLett.88.097601}%
  \BibitemOpen
  \bibfield  {author} {\bibinfo {author} {\bibfnamefont {J.}~\bibnamefont
  {Sigman}}, \bibinfo {author} {\bibfnamefont {D.~P.}\ \bibnamefont {Norton}},
  \bibinfo {author} {\bibfnamefont {H.~M.}\ \bibnamefont {Christen}}, \bibinfo
  {author} {\bibfnamefont {P.~H.}\ \bibnamefont {Fleming}},\ and\ \bibinfo
  {author} {\bibfnamefont {L.~A.}\ \bibnamefont {Boatner}},\ }\bibfield
  {title} {\bibinfo {title} {Antiferroelectric behavior in symmetric
  KNbO$_3$/KTaO$_3$ superlattices},\ }\href
  {https://doi.org/10.1103/PhysRevLett.88.097601} {\bibfield  {journal}
  {\bibinfo  {journal} {Phys. Rev. Lett.}\ }\textbf {\bibinfo {volume} {88}},\
  \bibinfo {pages} {097601} (\bibinfo {year} {2002})}\BibitemShut {NoStop}%
\bibitem [{\citenamefont {Kim}\ \emph {et~al.}(2002)\citenamefont {Kim},
  \citenamefont {Kim}, \citenamefont {Kim}, \citenamefont {Lee}, \citenamefont
  {Kim},\ and\ \citenamefont {Jung}}]{ApplPhysLett.80.3581}%
  \BibitemOpen
  \bibfield  {author} {\bibinfo {author} {\bibfnamefont {J.}~\bibnamefont
  {Kim}}, \bibinfo {author} {\bibfnamefont {Y.}~\bibnamefont {Kim}}, \bibinfo
  {author} {\bibfnamefont {Y.~S.}\ \bibnamefont {Kim}}, \bibinfo {author}
  {\bibfnamefont {J.}~\bibnamefont {Lee}}, \bibinfo {author} {\bibfnamefont
  {L.}~\bibnamefont {Kim}},\ and\ \bibinfo {author} {\bibfnamefont
  {D.}~\bibnamefont {Jung}},\ }\bibfield  {title} {\bibinfo {title} {Large
  nonlinear dielectric properties of artificial BaTiO$_3$/SrTiO$_3$
  superlattices},\ }\href {https://doi.org/10.1063/1.1477934} {\bibfield
  {journal} {\bibinfo  {journal} {Appl. Phys. Lett.}\ }\textbf {\bibinfo
  {volume} {80}},\ \bibinfo {pages} {3581} (\bibinfo {year}
  {2002})}\BibitemShut {NoStop}%
\bibitem [{\citenamefont {Kim}\ \emph {et~al.}(2003)\citenamefont {Kim},
  \citenamefont {Jung}, \citenamefont {Kim}, \citenamefont {Kim},\ and\
  \citenamefont {Lee}}]{ApplPhysLett.82.2118}%
  \BibitemOpen
  \bibfield  {author} {\bibinfo {author} {\bibfnamefont {L.}~\bibnamefont
  {Kim}}, \bibinfo {author} {\bibfnamefont {D.}~\bibnamefont {Jung}}, \bibinfo
  {author} {\bibfnamefont {J.}~\bibnamefont {Kim}}, \bibinfo {author}
  {\bibfnamefont {Y.~S.}\ \bibnamefont {Kim}},\ and\ \bibinfo {author}
  {\bibfnamefont {J.}~\bibnamefont {Lee}},\ }\bibfield  {title} {\bibinfo
  {title} {Strain manipulation in BaTiO$_3$/SrTiO$_3$ artificial lattice toward
  high dielectric constant and its nonlinearity},\ }\href
  {https://doi.org/10.1063/1.1565176} {\bibfield  {journal} {\bibinfo
  {journal} {Appl. Phys. Lett.}\ }\textbf {\bibinfo {volume} {82}},\ \bibinfo
  {pages} {2118} (\bibinfo {year} {2003})}\BibitemShut {NoStop}%
\bibitem [{\citenamefont {Jiang}\ \emph {et~al.}(2003)\citenamefont {Jiang},
  \citenamefont {Scott}, \citenamefont {Lu},\ and\ \citenamefont
  {Chen}}]{JApplPhys.93.1180}%
  \BibitemOpen
  \bibfield  {author} {\bibinfo {author} {\bibfnamefont {A.~Q.}\ \bibnamefont
  {Jiang}}, \bibinfo {author} {\bibfnamefont {J.~F.}\ \bibnamefont {Scott}},
  \bibinfo {author} {\bibfnamefont {H.}~\bibnamefont {Lu}},\ and\ \bibinfo
  {author} {\bibfnamefont {Z.}~\bibnamefont {Chen}},\ }\bibfield  {title}
  {\bibinfo {title} {Phase transitions and polarizations in epitaxial
  BaTiO$_3$/SrTiO$_3$ superlattices studied by second-harmonic generation},\
  }\href {https://doi.org/10.1063/1.1533094} {\bibfield  {journal} {\bibinfo
  {journal} {J. Appl. Phys.}\ }\textbf {\bibinfo {volume} {93}},\ \bibinfo
  {pages} {1180} (\bibinfo {year} {2003})}\BibitemShut {NoStop}%
\bibitem [{\citenamefont {Dawber}\ \emph {et~al.}(2005)\citenamefont {Dawber},
  \citenamefont {Lichtensteiger}, \citenamefont {Cantoni}, \citenamefont
  {Veithen}, \citenamefont {Ghosez}, \citenamefont {Johnston}, \citenamefont
  {Rabe},\ and\ \citenamefont {Triscone}}]{PhysRevLett.95.177601}%
  \BibitemOpen
  \bibfield  {author} {\bibinfo {author} {\bibfnamefont {M.}~\bibnamefont
  {Dawber}}, \bibinfo {author} {\bibfnamefont {C.}~\bibnamefont
  {Lichtensteiger}}, \bibinfo {author} {\bibfnamefont {M.}~\bibnamefont
  {Cantoni}}, \bibinfo {author} {\bibfnamefont {M.}~\bibnamefont {Veithen}},
  \bibinfo {author} {\bibfnamefont {P.}~\bibnamefont {Ghosez}}, \bibinfo
  {author} {\bibfnamefont {K.}~\bibnamefont {Johnston}}, \bibinfo {author}
  {\bibfnamefont {K.~M.}\ \bibnamefont {Rabe}},\ and\ \bibinfo {author}
  {\bibfnamefont {J.-M.}\ \bibnamefont {Triscone}},\ }\bibfield  {title}
  {\bibinfo {title} {Unusual behavior of the ferroelectric polarization in
  PbTiO$_3$/SrTiO$_3$ superlattices},\ }\href
  {https://doi.org/10.1103/PhysRevLett.95.177601} {\bibfield  {journal}
  {\bibinfo  {journal} {Phys. Rev. Lett.}\ }\textbf {\bibinfo {volume} {95}},\
  \bibinfo {pages} {177601} (\bibinfo {year} {2005})}\BibitemShut {NoStop}%
\bibitem [{\citenamefont {Tian}\ \emph {et~al.}(2006)\citenamefont {Tian},
  \citenamefont {Jiang}, \citenamefont {Pan}, \citenamefont {Haeni},
  \citenamefont {Li}, \citenamefont {Chen}, \citenamefont {Schlom},
  \citenamefont {Neaton}, \citenamefont {Rabe},\ and\ \citenamefont
  {Jia}}]{ApplPhysLett.89.092905}%
  \BibitemOpen
  \bibfield  {author} {\bibinfo {author} {\bibfnamefont {W.}~\bibnamefont
  {Tian}}, \bibinfo {author} {\bibfnamefont {J.~C.}\ \bibnamefont {Jiang}},
  \bibinfo {author} {\bibfnamefont {X.~Q.}\ \bibnamefont {Pan}}, \bibinfo
  {author} {\bibfnamefont {J.~H.}\ \bibnamefont {Haeni}}, \bibinfo {author}
  {\bibfnamefont {Y.~L.}\ \bibnamefont {Li}}, \bibinfo {author} {\bibfnamefont
  {L.~Q.}\ \bibnamefont {Chen}}, \bibinfo {author} {\bibfnamefont {D.~G.}\
  \bibnamefont {Schlom}}, \bibinfo {author} {\bibfnamefont {J.~B.}\
  \bibnamefont {Neaton}}, \bibinfo {author} {\bibfnamefont {K.~M.}\
  \bibnamefont {Rabe}},\ and\ \bibinfo {author} {\bibfnamefont {Q.~X.}\
  \bibnamefont {Jia}},\ }\bibfield  {title} {\bibinfo {title} {Structural
  evidence for enhanced polarization in a commensurate short-period
  BaTiO$_3$/SrTiO$_3$ superlattice},\ }\href
  {https://doi.org/10.1063/1.2335367} {\bibfield  {journal} {\bibinfo
  {journal} {Appl. Phys. Lett.}\ }\textbf {\bibinfo {volume} {89}},\ \bibinfo
  {pages} {092905} (\bibinfo {year} {2006})}\BibitemShut {NoStop}%
\bibitem [{\citenamefont {Tenne}\ \emph {et~al.}(2006)\citenamefont {Tenne},
  \citenamefont {Bruchhausen}, \citenamefont {Lanzillotti-Kimura},
  \citenamefont {Fainstein}, \citenamefont {Katiyar}, \citenamefont
  {Cantarero}, \citenamefont {Soukiassian}, \citenamefont {Vaithyanathan},
  \citenamefont {Haeni}, \citenamefont {Tian}, \citenamefont {Schlom},
  \citenamefont {Choi}, \citenamefont {Kim}, \citenamefont {Eom}, \citenamefont
  {Sun}, \citenamefont {Pan}, \citenamefont {Li}, \citenamefont {Chen},
  \citenamefont {Jia}, \citenamefont {Nakhmanson}, \citenamefont {Rabe},\ and\
  \citenamefont {Xi}}]{Science.313.1614}%
  \BibitemOpen
  \bibfield  {author} {\bibinfo {author} {\bibfnamefont {D.~A.}\ \bibnamefont
  {Tenne}}, \bibinfo {author} {\bibfnamefont {A.}~\bibnamefont {Bruchhausen}},
  \bibinfo {author} {\bibfnamefont {N.~D.}\ \bibnamefont {Lanzillotti-Kimura}},
  \bibinfo {author} {\bibfnamefont {A.}~\bibnamefont {Fainstein}}, \bibinfo
  {author} {\bibfnamefont {R.~S.}\ \bibnamefont {Katiyar}}, \bibinfo {author}
  {\bibfnamefont {A.}~\bibnamefont {Cantarero}}, \bibinfo {author}
  {\bibfnamefont {A.}~\bibnamefont {Soukiassian}}, \bibinfo {author}
  {\bibfnamefont {V.}~\bibnamefont {Vaithyanathan}}, \bibinfo {author}
  {\bibfnamefont {J.~H.}\ \bibnamefont {Haeni}}, \bibinfo {author}
  {\bibfnamefont {W.}~\bibnamefont {Tian}}, \bibinfo {author} {\bibfnamefont
  {D.~G.}\ \bibnamefont {Schlom}}, \bibinfo {author} {\bibfnamefont {K.~J.}\
  \bibnamefont {Choi}}, \bibinfo {author} {\bibfnamefont {D.~M.}\ \bibnamefont
  {Kim}}, \bibinfo {author} {\bibfnamefont {C.~B.}\ \bibnamefont {Eom}},
  \bibinfo {author} {\bibfnamefont {H.~P.}\ \bibnamefont {Sun}}, \bibinfo
  {author} {\bibfnamefont {X.~Q.}\ \bibnamefont {Pan}}, \bibinfo {author}
  {\bibfnamefont {Y.~L.}\ \bibnamefont {Li}}, \bibinfo {author} {\bibfnamefont
  {L.~Q.}\ \bibnamefont {Chen}}, \bibinfo {author} {\bibfnamefont {Q.~X.}\
  \bibnamefont {Jia}}, \bibinfo {author} {\bibfnamefont {S.~M.}\ \bibnamefont
  {Nakhmanson}}, \bibinfo {author} {\bibfnamefont {K.~M.}\ \bibnamefont
  {Rabe}},\ and\ \bibinfo {author} {\bibfnamefont {X.~X.}\ \bibnamefont {Xi}},\
  }\bibfield  {title} {\bibinfo {title} {Probing nanoscale ferroelectricity by
  ultraviolet Raman spectroscopy},\ }\href
  {https://doi.org/10.1126/science.1130306} {\bibfield  {journal} {\bibinfo
  {journal} {Science}\ }\textbf {\bibinfo {volume} {313}},\ \bibinfo {pages}
  {1614} (\bibinfo {year} {2006})}\BibitemShut {NoStop}%
\bibitem [{\citenamefont {Dawber}\ \emph {et~al.}(2007)\citenamefont {Dawber},
  \citenamefont {Stucki}, \citenamefont {Lichtensteiger}, \citenamefont
  {Gariglio}, \citenamefont {Ghosez},\ and\ \citenamefont
  {Triscone}}]{AdvMater.19.4153}%
  \BibitemOpen
  \bibfield  {author} {\bibinfo {author} {\bibfnamefont {M.}~\bibnamefont
  {Dawber}}, \bibinfo {author} {\bibfnamefont {N.}~\bibnamefont {Stucki}},
  \bibinfo {author} {\bibfnamefont {C.}~\bibnamefont {Lichtensteiger}},
  \bibinfo {author} {\bibfnamefont {S.}~\bibnamefont {Gariglio}}, \bibinfo
  {author} {\bibfnamefont {P.}~\bibnamefont {Ghosez}},\ and\ \bibinfo {author}
  {\bibfnamefont {J.-M.}\ \bibnamefont {Triscone}},\ }\bibfield  {title}
  {\bibinfo {title} {Tailoring the properties of artificially layered
  ferroelectric superlattices},\ }\href
  {https://doi.org/10.1002/adma.200700965} {\bibfield  {journal} {\bibinfo
  {journal} {Adv. Mater.}\ }\textbf {\bibinfo {volume} {19}},\ \bibinfo {pages}
  {4153} (\bibinfo {year} {2007})}\BibitemShut {NoStop}%
\bibitem [{\citenamefont {Zubko}\ \emph {et~al.}(2010)\citenamefont {Zubko},
  \citenamefont {Stucki}, \citenamefont {Lichtensteiger},\ and\ \citenamefont
  {Triscone}}]{PhysRevLett.104.187601}%
  \BibitemOpen
  \bibfield  {author} {\bibinfo {author} {\bibfnamefont {P.}~\bibnamefont
  {Zubko}}, \bibinfo {author} {\bibfnamefont {N.}~\bibnamefont {Stucki}},
  \bibinfo {author} {\bibfnamefont {C.}~\bibnamefont {Lichtensteiger}},\ and\
  \bibinfo {author} {\bibfnamefont {J.-M.}\ \bibnamefont {Triscone}},\
  }\bibfield  {title} {\bibinfo {title} {X-ray diffraction studies of
  180$^\circ$ ferroelectric domains in PbTiO$_3$/SrTiO$_3$ superlattices under
  an applied electric field},\ }\href
  {https://doi.org/10.1103/PhysRevLett.104.187601} {\bibfield  {journal}
  {\bibinfo  {journal} {Phys. Rev. Lett.}\ }\textbf {\bibinfo {volume} {104}},\
  \bibinfo {pages} {187601} (\bibinfo {year} {2010})}\BibitemShut {NoStop}%
\bibitem [{\citenamefont {Yuzyuk}(2012)}]{PhysSolidState.54.1026}%
  \BibitemOpen
  \bibfield  {author} {\bibinfo {author} {\bibfnamefont {Y.~I.}\ \bibnamefont
  {Yuzyuk}},\ }\bibfield  {title} {\bibinfo {title} {Raman scattering spectra
  of ceramics, films, and superlattices of ferroelectric perovskites: A
  review},\ }\href {https://doi.org/10.1134/S1063783412050502} {\bibfield
  {journal} {\bibinfo  {journal} {Phys. Solid State}\ }\textbf {\bibinfo
  {volume} {54}},\ \bibinfo {pages} {1026} (\bibinfo {year}
  {2012})}\BibitemShut {NoStop}%
\bibitem [{\citenamefont {Sinsheimer}\ \emph {et~al.}(2012)\citenamefont
  {Sinsheimer}, \citenamefont {Callori}, \citenamefont {Bein}, \citenamefont
  {Benkara}, \citenamefont {Daley}, \citenamefont {Coraor}, \citenamefont {Su},
  \citenamefont {Stephens},\ and\ \citenamefont
  {Dawber}}]{PhysRevLett.109.167601}%
  \BibitemOpen
  \bibfield  {author} {\bibinfo {author} {\bibfnamefont {J.}~\bibnamefont
  {Sinsheimer}}, \bibinfo {author} {\bibfnamefont {S.~J.}\ \bibnamefont
  {Callori}}, \bibinfo {author} {\bibfnamefont {B.}~\bibnamefont {Bein}},
  \bibinfo {author} {\bibfnamefont {Y.}~\bibnamefont {Benkara}}, \bibinfo
  {author} {\bibfnamefont {J.}~\bibnamefont {Daley}}, \bibinfo {author}
  {\bibfnamefont {J.}~\bibnamefont {Coraor}}, \bibinfo {author} {\bibfnamefont
  {D.}~\bibnamefont {Su}}, \bibinfo {author} {\bibfnamefont {P.~W.}\
  \bibnamefont {Stephens}},\ and\ \bibinfo {author} {\bibfnamefont
  {M.}~\bibnamefont {Dawber}},\ }\bibfield  {title} {\bibinfo {title}
  {Engineering polarization rotation in a ferroelectric superlattice},\ }\href
  {https://doi.org/10.1103/PhysRevLett.109.167601} {\bibfield  {journal}
  {\bibinfo  {journal} {Phys. Rev. Lett.}\ }\textbf {\bibinfo {volume} {109}},\
  \bibinfo {pages} {167601} (\bibinfo {year} {2012})}\BibitemShut {NoStop}%
\bibitem [{\citenamefont {Tikhonov}\ \emph {et~al.}(2015)\citenamefont
  {Tikhonov}, \citenamefont {Razumnaya}, \citenamefont {Maslova}, \citenamefont
  {Zakharchenko}, \citenamefont {Yuzyuk}, \citenamefont {Ortega}, \citenamefont
  {Kumar},\ and\ \citenamefont {Katiyar}}]{PhysSolidState.57.486}%
  \BibitemOpen
  \bibfield  {author} {\bibinfo {author} {\bibfnamefont {Y.~A.}\ \bibnamefont
  {Tikhonov}}, \bibinfo {author} {\bibfnamefont {A.~G.}\ \bibnamefont
  {Razumnaya}}, \bibinfo {author} {\bibfnamefont {O.~A.}\ \bibnamefont
  {Maslova}}, \bibinfo {author} {\bibfnamefont {I.~N.}\ \bibnamefont
  {Zakharchenko}}, \bibinfo {author} {\bibfnamefont {Y.~I.}\ \bibnamefont
  {Yuzyuk}}, \bibinfo {author} {\bibfnamefont {N.}~\bibnamefont {Ortega}},
  \bibinfo {author} {\bibfnamefont {A.}~\bibnamefont {Kumar}},\ and\ \bibinfo
  {author} {\bibfnamefont {R.~S.}\ \bibnamefont {Katiyar}},\ }\bibfield
  {title} {\bibinfo {title} {Phase transitions in two- and three-component
  perovskite superlattices},\ }\href
  {https://doi.org/10.1134/S1063783415030336} {\bibfield  {journal} {\bibinfo
  {journal} {Phys. Solid State}\ }\textbf {\bibinfo {volume} {57}},\ \bibinfo
  {pages} {486} (\bibinfo {year} {2015})}\BibitemShut {NoStop}%
\bibitem [{\citenamefont {Das}\ \emph {et~al.}(2019)\citenamefont {Das},
  \citenamefont {Tang}, \citenamefont {Hong}, \citenamefont {Gon\c{c}alves},
  \citenamefont {McCarter}, \citenamefont {Klewe}, \citenamefont {Nguyen},
  \citenamefont {G{\'o}mez-Ortiz}, \citenamefont {Shafer}, \citenamefont
  {Arenholz}, \citenamefont {Stoica}, \citenamefont {Hsu}, \citenamefont
  {Wang}, \citenamefont {Ophus}, \citenamefont {Liu}, \citenamefont {Nelson},
  \citenamefont {Saremi}, \citenamefont {Prasad}, \citenamefont {Mei},
  \citenamefont {Schlom}, \citenamefont {{\'I}{\~n}iguez}, \citenamefont
  {Garc{\'i}a-Fern{\'a}ndez}, \citenamefont {Muller}, \citenamefont {Chen},
  \citenamefont {Junquera}, \citenamefont {Martin},\ and\ \citenamefont
  {Ramesh}}]{Nature.568.368}%
  \BibitemOpen
  \bibfield  {author} {\bibinfo {author} {\bibfnamefont {S.}~\bibnamefont
  {Das}}, \bibinfo {author} {\bibfnamefont {Y.~L.}\ \bibnamefont {Tang}},
  \bibinfo {author} {\bibfnamefont {Z.}~\bibnamefont {Hong}}, \bibinfo {author}
  {\bibfnamefont {M.~A.~P.}\ \bibnamefont {Gon\c{c}alves}}, \bibinfo {author}
  {\bibfnamefont {M.~R.}\ \bibnamefont {McCarter}}, \bibinfo {author}
  {\bibfnamefont {C.}~\bibnamefont {Klewe}}, \bibinfo {author} {\bibfnamefont
  {K.~X.}\ \bibnamefont {Nguyen}}, \bibinfo {author} {\bibfnamefont
  {F.}~\bibnamefont {G{\'o}mez-Ortiz}}, \bibinfo {author} {\bibfnamefont
  {P.}~\bibnamefont {Shafer}}, \bibinfo {author} {\bibfnamefont
  {E.}~\bibnamefont {Arenholz}}, \bibinfo {author} {\bibfnamefont {V.~A.}\
  \bibnamefont {Stoica}}, \bibinfo {author} {\bibfnamefont {S.-L.}\
  \bibnamefont {Hsu}}, \bibinfo {author} {\bibfnamefont {B.}~\bibnamefont
  {Wang}}, \bibinfo {author} {\bibfnamefont {C.}~\bibnamefont {Ophus}},
  \bibinfo {author} {\bibfnamefont {J.~F.}\ \bibnamefont {Liu}}, \bibinfo
  {author} {\bibfnamefont {C.~T.}\ \bibnamefont {Nelson}}, \bibinfo {author}
  {\bibfnamefont {S.}~\bibnamefont {Saremi}}, \bibinfo {author} {\bibfnamefont
  {B.}~\bibnamefont {Prasad}}, \bibinfo {author} {\bibfnamefont {A.~B.}\
  \bibnamefont {Mei}}, \bibinfo {author} {\bibfnamefont {D.~G.}\ \bibnamefont
  {Schlom}}, \bibinfo {author} {\bibfnamefont {J.}~\bibnamefont
  {{\'I}{\~n}iguez}}, \bibinfo {author} {\bibfnamefont {P.}~\bibnamefont
  {Garc{\'i}a-Fern{\'a}ndez}}, \bibinfo {author} {\bibfnamefont {D.~A.}\
  \bibnamefont {Muller}}, \bibinfo {author} {\bibfnamefont {L.~Q.}\
  \bibnamefont {Chen}}, \bibinfo {author} {\bibfnamefont {J.}~\bibnamefont
  {Junquera}}, \bibinfo {author} {\bibfnamefont {L.~W.}\ \bibnamefont
  {Martin}},\ and\ \bibinfo {author} {\bibfnamefont {R.}~\bibnamefont
  {Ramesh}},\ }\bibfield  {title} {\bibinfo {title} {Observation of
  room-temperature polar skyrmions},\ }\href
  {https://doi.org/10.1038/s41586-019-1092-8} {\bibfield  {journal} {\bibinfo
  {journal} {Nature}\ }\textbf {\bibinfo {volume} {568}},\ \bibinfo {pages}
  {368} (\bibinfo {year} {2019})}\BibitemShut {NoStop}%
\bibitem [{\citenamefont {Sidorkin}\ \emph {et~al.}(2019)\citenamefont
  {Sidorkin}, \citenamefont {Nesterenko}, \citenamefont {Gagou}, \citenamefont
  {Saint-Gregoire}, \citenamefont {Vorotnikov}, \citenamefont {Pakhomov},\ and\
  \citenamefont {Popravko}}]{SciRep.9.18948}%
  \BibitemOpen
  \bibfield  {author} {\bibinfo {author} {\bibfnamefont {A.~S.}\ \bibnamefont
  {Sidorkin}}, \bibinfo {author} {\bibfnamefont {L.~P.}\ \bibnamefont
  {Nesterenko}}, \bibinfo {author} {\bibfnamefont {Y.}~\bibnamefont {Gagou}},
  \bibinfo {author} {\bibfnamefont {P.}~\bibnamefont {Saint-Gregoire}},
  \bibinfo {author} {\bibfnamefont {E.~V.}\ \bibnamefont {Vorotnikov}},
  \bibinfo {author} {\bibfnamefont {A.~Y.}\ \bibnamefont {Pakhomov}},\ and\
  \bibinfo {author} {\bibfnamefont {N.~G.}\ \bibnamefont {Popravko}},\
  }\bibfield  {title} {\bibinfo {title} {Repolarization of ferroelectric
  superlattices BaZrO$_3$/BaTiO$_3$},\ }\href
  {https://doi.org/10.1038/s41598-019-55475-2} {\bibfield  {journal} {\bibinfo
  {journal} {Sci. Rep.}\ }\textbf {\bibinfo {volume} {9}},\ \bibinfo {pages}
  {18948} (\bibinfo {year} {2019})}\BibitemShut {NoStop}%
\bibitem [{\citenamefont {S\'aghi-Szab\'o}\ \emph {et~al.}(1999)\citenamefont
  {S\'aghi-Szab\'o}, \citenamefont {Cohen},\ and\ \citenamefont
  {Krakauer}}]{PhysRevB.59.12771}%
  \BibitemOpen
  \bibfield  {author} {\bibinfo {author} {\bibfnamefont {G.}~\bibnamefont
  {S\'aghi-Szab\'o}}, \bibinfo {author} {\bibfnamefont {R.~E.}\ \bibnamefont
  {Cohen}},\ and\ \bibinfo {author} {\bibfnamefont {H.}~\bibnamefont
  {Krakauer}},\ }\bibfield  {title} {\bibinfo {title} {First-principles study
  of piezoelectricity in tetragonal PbTiO$_3$ and
  PbZr$_{1/2}$Ti$_{1/2}$O$_3$},\ }\href
  {https://doi.org/10.1103/PhysRevB.59.12771} {\bibfield  {journal} {\bibinfo
  {journal} {Phys. Rev. B}\ }\textbf {\bibinfo {volume} {59}},\ \bibinfo
  {pages} {12771} (\bibinfo {year} {1999})}\BibitemShut {NoStop}%
\bibitem [{\citenamefont {Sai}\ \emph {et~al.}(2000)\citenamefont {Sai},
  \citenamefont {Meyer},\ and\ \citenamefont
  {Vanderbilt}}]{PhysRevLett.84.5636}%
  \BibitemOpen
  \bibfield  {author} {\bibinfo {author} {\bibfnamefont {N.}~\bibnamefont
  {Sai}}, \bibinfo {author} {\bibfnamefont {B.}~\bibnamefont {Meyer}},\ and\
  \bibinfo {author} {\bibfnamefont {D.}~\bibnamefont {Vanderbilt}},\ }\bibfield
   {title} {\bibinfo {title} {Compositional inversion symmetry breaking in
  ferroelectric perovskites},\ }\href
  {https://doi.org/10.1103/PhysRevLett.84.5636} {\bibfield  {journal} {\bibinfo
   {journal} {Phys. Rev. Lett.}\ }\textbf {\bibinfo {volume} {84}},\ \bibinfo
  {pages} {5636} (\bibinfo {year} {2000})}\BibitemShut {NoStop}%
\bibitem [{\citenamefont {Neaton}\ and\ \citenamefont
  {Rabe}(2003)}]{ApplPhysLett.82.1586}%
  \BibitemOpen
  \bibfield  {author} {\bibinfo {author} {\bibfnamefont {J.~B.}\ \bibnamefont
  {Neaton}}\ and\ \bibinfo {author} {\bibfnamefont {K.~M.}\ \bibnamefont
  {Rabe}},\ }\bibfield  {title} {\bibinfo {title} {Theory of polarization
  enhancement in epitaxial BaTiO$_3$/SrTiO$_3$ superlattices},\ }\href
  {https://doi.org/10.1063/1.1559651} {\bibfield  {journal} {\bibinfo
  {journal} {Appl. Phys. Lett.}\ }\textbf {\bibinfo {volume} {82}},\ \bibinfo
  {pages} {1586} (\bibinfo {year} {2003})}\BibitemShut {NoStop}%
\bibitem [{\citenamefont {Bungaro}\ and\ \citenamefont
  {Rabe}(2004)}]{PhysRevB.69.184101}%
  \BibitemOpen
  \bibfield  {author} {\bibinfo {author} {\bibfnamefont {C.}~\bibnamefont
  {Bungaro}}\ and\ \bibinfo {author} {\bibfnamefont {K.~M.}\ \bibnamefont
  {Rabe}},\ }\bibfield  {title} {\bibinfo {title} {Epitaxially strained
  [001]-(PbTiO$_3$)$_1$(PbZrO$_3$)$_1$ superlattice and PbTiO$_3$ from first
  principles},\ }\href {https://doi.org/10.1103/PhysRevB.69.184101} {\bibfield
  {journal} {\bibinfo  {journal} {Phys. Rev. B}\ }\textbf {\bibinfo {volume}
  {69}},\ \bibinfo {pages} {184101} (\bibinfo {year} {2004})}\BibitemShut
  {NoStop}%
\bibitem [{\citenamefont {Lee}\ \emph {et~al.}(2005)\citenamefont {Lee},
  \citenamefont {Christen}, \citenamefont {Chisholm}, \citenamefont {Rouleau},\
  and\ \citenamefont {Lowndes}}]{Nature.433.395}%
  \BibitemOpen
  \bibfield  {author} {\bibinfo {author} {\bibfnamefont {H.~N.}\ \bibnamefont
  {Lee}}, \bibinfo {author} {\bibfnamefont {H.~M.}\ \bibnamefont {Christen}},
  \bibinfo {author} {\bibfnamefont {M.~F.}\ \bibnamefont {Chisholm}}, \bibinfo
  {author} {\bibfnamefont {C.~M.}\ \bibnamefont {Rouleau}},\ and\ \bibinfo
  {author} {\bibfnamefont {D.~H.}\ \bibnamefont {Lowndes}},\ }\bibfield
  {title} {\bibinfo {title} {Strong polarization enhancement in asymmetric
  three-component ferroelectric superlattices},\ }\href
  {https://doi.org/10.1038/nature03261} {\bibfield  {journal} {\bibinfo
  {journal} {Nature}\ }\textbf {\bibinfo {volume} {433}},\ \bibinfo {pages}
  {395} (\bibinfo {year} {2005})}\BibitemShut {NoStop}%
\bibitem [{\citenamefont {Johnston}\ \emph {et~al.}(2005)\citenamefont
  {Johnston}, \citenamefont {Huang}, \citenamefont {Neaton},\ and\
  \citenamefont {Rabe}}]{PhysRevB.71.100103}%
  \BibitemOpen
  \bibfield  {author} {\bibinfo {author} {\bibfnamefont {K.}~\bibnamefont
  {Johnston}}, \bibinfo {author} {\bibfnamefont {X.}~\bibnamefont {Huang}},
  \bibinfo {author} {\bibfnamefont {J.~B.}\ \bibnamefont {Neaton}},\ and\
  \bibinfo {author} {\bibfnamefont {K.~M.}\ \bibnamefont {Rabe}},\ }\bibfield
  {title} {\bibinfo {title} {First-principles study of symmetry lowering and
  polarization in BaTiO$_3$/SrTiO$_3$ superlattices with in-plane expansion},\
  }\href {https://doi.org/10.1103/PhysRevB.71.100103} {\bibfield  {journal}
  {\bibinfo  {journal} {Phys. Rev. B}\ }\textbf {\bibinfo {volume} {71}},\
  \bibinfo {pages} {100103} (\bibinfo {year} {2005})}\BibitemShut {NoStop}%
\bibitem [{\citenamefont {Kim}\ \emph {et~al.}(2005)\citenamefont {Kim},
  \citenamefont {Kim}, \citenamefont {Jung}, \citenamefont {Lee},\ and\
  \citenamefont {Waghmare}}]{ApplPhysLett.87.052903}%
  \BibitemOpen
  \bibfield  {author} {\bibinfo {author} {\bibfnamefont {L.}~\bibnamefont
  {Kim}}, \bibinfo {author} {\bibfnamefont {J.}~\bibnamefont {Kim}}, \bibinfo
  {author} {\bibfnamefont {D.}~\bibnamefont {Jung}}, \bibinfo {author}
  {\bibfnamefont {J.}~\bibnamefont {Lee}},\ and\ \bibinfo {author}
  {\bibfnamefont {U.~V.}\ \bibnamefont {Waghmare}},\ }\bibfield  {title}
  {\bibinfo {title} {Polarization of strained BaTiO$_3$/SrTiO$_3$ artificial
  superlattice: First-principles study},\ }\href
  {https://doi.org/10.1063/1.2006216} {\bibfield  {journal} {\bibinfo
  {journal} {Appl. Phys. Lett.}\ }\textbf {\bibinfo {volume} {87}},\ \bibinfo
  {pages} {052903} (\bibinfo {year} {2005})}\BibitemShut {NoStop}%
\bibitem [{\citenamefont {Nakhmanson}\ \emph {et~al.}(2005)\citenamefont
  {Nakhmanson}, \citenamefont {Rabe},\ and\ \citenamefont
  {Vanderbilt}}]{ApplPhysLett.87.102906}%
  \BibitemOpen
  \bibfield  {author} {\bibinfo {author} {\bibfnamefont {S.~M.}\ \bibnamefont
  {Nakhmanson}}, \bibinfo {author} {\bibfnamefont {K.~M.}\ \bibnamefont
  {Rabe}},\ and\ \bibinfo {author} {\bibfnamefont {D.}~\bibnamefont
  {Vanderbilt}},\ }\bibfield  {title} {\bibinfo {title} {Polarization
  enhancement in two- and three-component ferroelectric superlattices},\ }\href
  {https://doi.org/10.1063/1.2042630} {\bibfield  {journal} {\bibinfo
  {journal} {Appl. Phys. Lett.}\ }\textbf {\bibinfo {volume} {87}},\ \bibinfo
  {pages} {102906} (\bibinfo {year} {2005})}\BibitemShut {NoStop}%
\bibitem [{\citenamefont {Ghosez}\ and\ \citenamefont
  {Junquera}(2006)}]{HandbookChap134}%
  \BibitemOpen
  \bibfield  {author} {\bibinfo {author} {\bibfnamefont {P.}~\bibnamefont
  {Ghosez}}\ and\ \bibinfo {author} {\bibfnamefont {J.}~\bibnamefont
  {Junquera}},\ }\bibfield  {title} {\bibinfo {title} {First-principle modeling
  of ferroelectric oxide nanostructures},\ }in\ \href@noop {} {\emph {\bibinfo
  {booktitle} {Handbook of Theoretical and Computational Nanotechnology}}},\
  Vol.~\bibinfo {volume} {9},\ \bibinfo {editor} {edited by\ \bibinfo {editor}
  {\bibfnamefont {M.}~\bibnamefont {Rieth}}\ and\ \bibinfo {editor}
  {\bibfnamefont {W.}~\bibnamefont {Schommers}}}\ (\bibinfo  {publisher}
  {American Scientific Publishers},\ \bibinfo {year} {2006})\ pp.\ \bibinfo
  {pages} {623--728}\BibitemShut {NoStop}%
\bibitem [{\citenamefont {Lisenkov}\ and\ \citenamefont
  {Bellaiche}(2007)}]{PhysRevB.76.020102}%
  \BibitemOpen
  \bibfield  {author} {\bibinfo {author} {\bibfnamefont {S.}~\bibnamefont
  {Lisenkov}}\ and\ \bibinfo {author} {\bibfnamefont {L.}~\bibnamefont
  {Bellaiche}},\ }\bibfield  {title} {\bibinfo {title} {Phase diagrams of
  BaTiO$_3$/SrTiO$_3$ superlattices from first principles},\ }\href
  {https://doi.org/10.1103/PhysRevB.76.020102} {\bibfield  {journal} {\bibinfo
  {journal} {Phys. Rev. B}\ }\textbf {\bibinfo {volume} {76}},\ \bibinfo
  {pages} {020102} (\bibinfo {year} {2007})}\BibitemShut {NoStop}%
\bibitem [{\citenamefont {Li}\ \emph {et~al.}(2007)\citenamefont {Li},
  \citenamefont {Hu}, \citenamefont {Tenne}, \citenamefont {Soukiassian},
  \citenamefont {Schlom}, \citenamefont {Xi}, \citenamefont {Choi},
  \citenamefont {Eom}, \citenamefont {Saxena}, \citenamefont {Lookman},
  \citenamefont {Jia},\ and\ \citenamefont {Chen}}]{ApplPhysLett.91.112914}%
  \BibitemOpen
  \bibfield  {author} {\bibinfo {author} {\bibfnamefont {Y.~L.}\ \bibnamefont
  {Li}}, \bibinfo {author} {\bibfnamefont {S.~Y.}\ \bibnamefont {Hu}}, \bibinfo
  {author} {\bibfnamefont {D.}~\bibnamefont {Tenne}}, \bibinfo {author}
  {\bibfnamefont {A.}~\bibnamefont {Soukiassian}}, \bibinfo {author}
  {\bibfnamefont {D.~G.}\ \bibnamefont {Schlom}}, \bibinfo {author}
  {\bibfnamefont {X.~X.}\ \bibnamefont {Xi}}, \bibinfo {author} {\bibfnamefont
  {K.~J.}\ \bibnamefont {Choi}}, \bibinfo {author} {\bibfnamefont {C.~B.}\
  \bibnamefont {Eom}}, \bibinfo {author} {\bibfnamefont {A.}~\bibnamefont
  {Saxena}}, \bibinfo {author} {\bibfnamefont {T.}~\bibnamefont {Lookman}},
  \bibinfo {author} {\bibfnamefont {Q.~X.}\ \bibnamefont {Jia}},\ and\ \bibinfo
  {author} {\bibfnamefont {L.~Q.}\ \bibnamefont {Chen}},\ }\bibfield  {title}
  {\bibinfo {title} {Prediction of ferroelectricity in BaTiO$_3$/SrTiO$_3$
  superlattices with domains},\ }\href {https://doi.org/10.1063/1.2785121}
  {\bibfield  {journal} {\bibinfo  {journal} {Appl. Phys. Lett.}\ }\textbf
  {\bibinfo {volume} {91}},\ \bibinfo {pages} {112914} (\bibinfo {year}
  {2007})}\BibitemShut {NoStop}%
\bibitem [{\citenamefont {Bousquet}\ \emph {et~al.}(2008)\citenamefont
  {Bousquet}, \citenamefont {Dawber}, \citenamefont {Stucki}, \citenamefont
  {Lichtensteiger}, \citenamefont {Hermet}, \citenamefont {Gariglio},
  \citenamefont {Triscone},\ and\ \citenamefont {Ghosez}}]{Nature.452.732}%
  \BibitemOpen
  \bibfield  {author} {\bibinfo {author} {\bibfnamefont {E.}~\bibnamefont
  {Bousquet}}, \bibinfo {author} {\bibfnamefont {M.}~\bibnamefont {Dawber}},
  \bibinfo {author} {\bibfnamefont {N.}~\bibnamefont {Stucki}}, \bibinfo
  {author} {\bibfnamefont {C.}~\bibnamefont {Lichtensteiger}}, \bibinfo
  {author} {\bibfnamefont {P.}~\bibnamefont {Hermet}}, \bibinfo {author}
  {\bibfnamefont {S.}~\bibnamefont {Gariglio}}, \bibinfo {author}
  {\bibfnamefont {J.-M.}\ \bibnamefont {Triscone}},\ and\ \bibinfo {author}
  {\bibfnamefont {P.}~\bibnamefont {Ghosez}},\ }\bibfield  {title} {\bibinfo
  {title} {Improper ferroelectricity in perovskite oxide artificial
  superlattices},\ }\href {https://doi.org/10.1038/nature06817} {\bibfield
  {journal} {\bibinfo  {journal} {Nature}\ }\textbf {\bibinfo {volume} {452}},\
  \bibinfo {pages} {732} (\bibinfo {year} {2008})}\BibitemShut {NoStop}%
\bibitem [{\citenamefont {Wu}\ \emph {et~al.}(2008)\citenamefont {Wu},
  \citenamefont {Stengel}, \citenamefont {Rabe},\ and\ \citenamefont
  {Vanderbilt}}]{PhysRevLett.101.087601}%
  \BibitemOpen
  \bibfield  {author} {\bibinfo {author} {\bibfnamefont {X.}~\bibnamefont
  {Wu}}, \bibinfo {author} {\bibfnamefont {M.}~\bibnamefont {Stengel}},
  \bibinfo {author} {\bibfnamefont {K.~M.}\ \bibnamefont {Rabe}},\ and\
  \bibinfo {author} {\bibfnamefont {D.}~\bibnamefont {Vanderbilt}},\ }\bibfield
   {title} {\bibinfo {title} {Predicting polarization and nonlinear dielectric
  response of arbitrary perovskite superlattice sequences},\ }\href
  {https://doi.org/10.1103/PhysRevLett.101.087601} {\bibfield  {journal}
  {\bibinfo  {journal} {Phys. Rev. Lett.}\ }\textbf {\bibinfo {volume} {101}},\
  \bibinfo {pages} {087601} (\bibinfo {year} {2008})}\BibitemShut {NoStop}%
\bibitem [{\citenamefont {Lisenkov}\ \emph {et~al.}(2009)\citenamefont
  {Lisenkov}, \citenamefont {Ponomareva},\ and\ \citenamefont
  {Bellaiche}}]{PhysRevB.79.024101}%
  \BibitemOpen
  \bibfield  {author} {\bibinfo {author} {\bibfnamefont {S.}~\bibnamefont
  {Lisenkov}}, \bibinfo {author} {\bibfnamefont {I.}~\bibnamefont
  {Ponomareva}},\ and\ \bibinfo {author} {\bibfnamefont {L.}~\bibnamefont
  {Bellaiche}},\ }\bibfield  {title} {\bibinfo {title} {Unusual static and
  dynamical characteristics of domain evolution in ferroelectric
  superlattices},\ }\href {https://doi.org/10.1103/PhysRevB.79.024101}
  {\bibfield  {journal} {\bibinfo  {journal} {Phys. Rev. B}\ }\textbf {\bibinfo
  {volume} {79}},\ \bibinfo {pages} {024101} (\bibinfo {year}
  {2009})}\BibitemShut {NoStop}%
\bibitem [{\citenamefont
  {Lebedev}(2009{\natexlab{a}})}]{PhysSolidState.51.2324}%
  \BibitemOpen
  \bibfield  {author} {\bibinfo {author} {\bibfnamefont {A.~I.}\ \bibnamefont
  {Lebedev}},\ }\bibfield  {title} {\bibinfo {title} {Ab initio studies of
  dielectric, piezoelectric, and elastic properties of BaTiO$_3$/SrTiO$_3$
  ferroelectric superlattices},\ }\href
  {https://doi.org/10.1134/S1063783409110225} {\bibfield  {journal} {\bibinfo
  {journal} {Phys. Solid State}\ }\textbf {\bibinfo {volume} {51}},\ \bibinfo
  {pages} {2324} (\bibinfo {year} {2009}{\natexlab{a}})}\BibitemShut {NoStop}%
\bibitem [{\citenamefont {Lebedev}(2010)}]{PhysSolidState.52.1448}%
  \BibitemOpen
  \bibfield  {author} {\bibinfo {author} {\bibfnamefont {A.~I.}\ \bibnamefont
  {Lebedev}},\ }\bibfield  {title} {\bibinfo {title} {Ground state and
  properties of ferroelectric superlattices based on crystals of the perovskite
  family},\ }\href {https://doi.org/10.1134/S1063783410070218} {\bibfield
  {journal} {\bibinfo  {journal} {Phys. Solid State}\ }\textbf {\bibinfo
  {volume} {52}},\ \bibinfo {pages} {1448} (\bibinfo {year}
  {2010})}\BibitemShut {NoStop}%
\bibitem [{\citenamefont {Wu}\ \emph {et~al.}(2011)\citenamefont {Wu},
  \citenamefont {Rabe},\ and\ \citenamefont {Vanderbilt}}]{PhysRevB.83.020104}%
  \BibitemOpen
  \bibfield  {author} {\bibinfo {author} {\bibfnamefont {X.}~\bibnamefont
  {Wu}}, \bibinfo {author} {\bibfnamefont {K.~M.}\ \bibnamefont {Rabe}},\ and\
  \bibinfo {author} {\bibfnamefont {D.}~\bibnamefont {Vanderbilt}},\ }\bibfield
   {title} {\bibinfo {title} {Interfacial enhancement of ferroelectricity in
  CaTiO$_3$/BaTiO$_3$ superlattices},\ }\href
  {https://doi.org/10.1103/PhysRevB.83.020104} {\bibfield  {journal} {\bibinfo
  {journal} {Phys. Rev. B}\ }\textbf {\bibinfo {volume} {83}},\ \bibinfo
  {pages} {020104} (\bibinfo {year} {2011})}\BibitemShut {NoStop}%
\bibitem [{\citenamefont {Aguado-Puente}\ \emph {et~al.}(2011)\citenamefont
  {Aguado-Puente}, \citenamefont {Garc{\'i}a-Fern{\'a}ndez},\ and\
  \citenamefont {Junquera}}]{PhysRevLett.107.217601}%
  \BibitemOpen
  \bibfield  {author} {\bibinfo {author} {\bibfnamefont {P.}~\bibnamefont
  {Aguado-Puente}}, \bibinfo {author} {\bibfnamefont {P.}~\bibnamefont
  {Garc{\'i}a-Fern{\'a}ndez}},\ and\ \bibinfo {author} {\bibfnamefont
  {J.}~\bibnamefont {Junquera}},\ }\bibfield  {title} {\bibinfo {title}
  {Interplay of couplings between antiferrodistortive, ferroelectric, and
  strain degrees of freedom in monodomain PbTiO$_3$/SrTiO$_3$ superlattices},\
  }\href {https://doi.org/10.1103/PhysRevLett.107.217601} {\bibfield  {journal}
  {\bibinfo  {journal} {Phys. Rev. Lett.}\ }\textbf {\bibinfo {volume} {107}},\
  \bibinfo {pages} {217601} (\bibinfo {year} {2011})}\BibitemShut {NoStop}%
\bibitem [{\citenamefont {Lebedev}(2012)}]{PhysStatusSolidiB.249.789}%
  \BibitemOpen
  \bibfield  {author} {\bibinfo {author} {\bibfnamefont {A.~I.}\ \bibnamefont
  {Lebedev}},\ }\bibfield  {title} {\bibinfo {title} {Ground-state structure of
  KNbO$_3$/KTaO$_3$ superlattices: Array of nearly independent
  ferroelectrically ordered planes},\ }\href
  {https://doi.org/10.1002/pssb.201147350} {\bibfield  {journal} {\bibinfo
  {journal} {Phys. Status Solidi B}\ }\textbf {\bibinfo {volume} {249}},\
  \bibinfo {pages} {789} (\bibinfo {year} {2012})}\BibitemShut {NoStop}%
\bibitem [{\citenamefont {Aguado-Puente}\ and\ \citenamefont
  {Junquera}(2012)}]{PhysRevB.85.184105}%
  \BibitemOpen
  \bibfield  {author} {\bibinfo {author} {\bibfnamefont {P.}~\bibnamefont
  {Aguado-Puente}}\ and\ \bibinfo {author} {\bibfnamefont {J.}~\bibnamefont
  {Junquera}},\ }\bibfield  {title} {\bibinfo {title} {Structural and energetic
  properties of domains in PbTiO$_3$/SrTiO$_3$ superlattices from first
  principles},\ }\href {https://doi.org/10.1103/PhysRevB.85.184105} {\bibfield
  {journal} {\bibinfo  {journal} {Phys. Rev. B}\ }\textbf {\bibinfo {volume}
  {85}},\ \bibinfo {pages} {184105} (\bibinfo {year} {2012})}\BibitemShut
  {NoStop}%
\bibitem [{\citenamefont {Lebedev}(2013)}]{PhysSolidState.55.1198}%
  \BibitemOpen
  \bibfield  {author} {\bibinfo {author} {\bibfnamefont {A.~I.}\ \bibnamefont
  {Lebedev}},\ }\bibfield  {title} {\bibinfo {title} {Properties of
  BaTiO$_3$/BaZrO$_3$ ferroelectric superlattices with competing
  instabilities},\ }\href {https://doi.org/10.1134/S1063783413060218}
  {\bibfield  {journal} {\bibinfo  {journal} {Phys. Solid State}\ }\textbf
  {\bibinfo {volume} {55}},\ \bibinfo {pages} {1198} (\bibinfo {year}
  {2013})}\BibitemShut {NoStop}%
\bibitem [{\citenamefont {Lu}\ \emph {et~al.}(2014)\citenamefont {Lu},
  \citenamefont {Gong},\ and\ \citenamefont {Xiang}}]{ComputMaterSci.91.310}%
  \BibitemOpen
  \bibfield  {author} {\bibinfo {author} {\bibfnamefont {X.~Z.}\ \bibnamefont
  {Lu}}, \bibinfo {author} {\bibfnamefont {X.~G.}\ \bibnamefont {Gong}},\ and\
  \bibinfo {author} {\bibfnamefont {H.~J.}\ \bibnamefont {Xiang}},\ }\bibfield
  {title} {\bibinfo {title} {Polarization enhancement in perovskite
  superlattices by oxygen octahedral tilts},\ }\href
  {https://doi.org/10.1016/j.commatsci.2014.05.003} {\bibfield  {journal}
  {\bibinfo  {journal} {Comput. Mater. Sci.}\ }\textbf {\bibinfo {volume}
  {91}},\ \bibinfo {pages} {310} (\bibinfo {year} {2014})}\BibitemShut
  {NoStop}%
\bibitem [{\citenamefont {Zubko}\ \emph {et~al.}(2016)\citenamefont {Zubko},
  \citenamefont {Wojde{\l}}, \citenamefont {Hadjimichael}, \citenamefont
  {Fernandez-Pena}, \citenamefont {Sen{\'e}}, \citenamefont {Luk'yanchuk},
  \citenamefont {Triscone},\ and\ \citenamefont
  {{\'I}{\~n}iguez}}]{Nature.534.524}%
  \BibitemOpen
  \bibfield  {author} {\bibinfo {author} {\bibfnamefont {P.}~\bibnamefont
  {Zubko}}, \bibinfo {author} {\bibfnamefont {J.~C.}\ \bibnamefont
  {Wojde{\l}}}, \bibinfo {author} {\bibfnamefont {M.}~\bibnamefont
  {Hadjimichael}}, \bibinfo {author} {\bibfnamefont {S.}~\bibnamefont
  {Fernandez-Pena}}, \bibinfo {author} {\bibfnamefont {A.}~\bibnamefont
  {Sen{\'e}}}, \bibinfo {author} {\bibfnamefont {I.}~\bibnamefont
  {Luk'yanchuk}}, \bibinfo {author} {\bibfnamefont {J.-M.}\ \bibnamefont
  {Triscone}},\ and\ \bibinfo {author} {\bibfnamefont {J.}~\bibnamefont
  {{\'I}{\~n}iguez}},\ }\bibfield  {title} {\bibinfo {title} {Negative
  capacitance in multidomain ferroelectric superlattices},\ }\href
  {https://doi.org/10.1038/nature17659} {\bibfield  {journal} {\bibinfo
  {journal} {Nature}\ }\textbf {\bibinfo {volume} {534}},\ \bibinfo {pages}
  {524} (\bibinfo {year} {2016})}\BibitemShut {NoStop}%
\bibitem [{\citenamefont {Hong}\ \emph {et~al.}(2017)\citenamefont {Hong},
  \citenamefont {Damodaran}, \citenamefont {Xue}, \citenamefont {Hsu},
  \citenamefont {Britson}, \citenamefont {Yadav}, \citenamefont {Nelson},
  \citenamefont {Wang}, \citenamefont {Scott}, \citenamefont {Martin},
  \citenamefont {Ramesh},\ and\ \citenamefont {Chen}}]{NanoLett.17.2246}%
  \BibitemOpen
  \bibfield  {author} {\bibinfo {author} {\bibfnamefont {Z.}~\bibnamefont
  {Hong}}, \bibinfo {author} {\bibfnamefont {A.~R.}\ \bibnamefont {Damodaran}},
  \bibinfo {author} {\bibfnamefont {F.}~\bibnamefont {Xue}}, \bibinfo {author}
  {\bibfnamefont {S.-L.}\ \bibnamefont {Hsu}}, \bibinfo {author} {\bibfnamefont
  {J.}~\bibnamefont {Britson}}, \bibinfo {author} {\bibfnamefont {A.~K.}\
  \bibnamefont {Yadav}}, \bibinfo {author} {\bibfnamefont {C.~T.}\ \bibnamefont
  {Nelson}}, \bibinfo {author} {\bibfnamefont {J.-J.}\ \bibnamefont {Wang}},
  \bibinfo {author} {\bibfnamefont {J.~F.}\ \bibnamefont {Scott}}, \bibinfo
  {author} {\bibfnamefont {L.~W.}\ \bibnamefont {Martin}}, \bibinfo {author}
  {\bibfnamefont {R.}~\bibnamefont {Ramesh}},\ and\ \bibinfo {author}
  {\bibfnamefont {L.-Q.}\ \bibnamefont {Chen}},\ }\bibfield  {title} {\bibinfo
  {title} {Stability of polar vortex lattice in ferroelectric superlattices},\
  }\href {https://doi.org/10.1021/acs.nanolett.6b04875} {\bibfield  {journal}
  {\bibinfo  {journal} {Nano Lett.}\ }\textbf {\bibinfo {volume} {17}},\
  \bibinfo {pages} {2246} (\bibinfo {year} {2017})}\BibitemShut {NoStop}%
\bibitem [{\citenamefont {Hong}\ and\ \citenamefont
  {Chen}(2018)}]{ActaMater.152.155}%
  \BibitemOpen
  \bibfield  {author} {\bibinfo {author} {\bibfnamefont {Z.}~\bibnamefont
  {Hong}}\ and\ \bibinfo {author} {\bibfnamefont {L.-Q.}\ \bibnamefont
  {Chen}},\ }\bibfield  {title} {\bibinfo {title} {Blowing polar skyrmion
  bubbles in oxide superlattices},\ }\href
  {https://doi.org/10.1016/j.actamat.2018.04.022} {\bibfield  {journal}
  {\bibinfo  {journal} {Acta Mater.}\ }\textbf {\bibinfo {volume} {152}},\
  \bibinfo {pages} {155} (\bibinfo {year} {2018})}\BibitemShut {NoStop}%
\bibitem [{\citenamefont {Ohtomo}\ and\ \citenamefont
  {Hwang}(2004)}]{Nature.427.423}%
  \BibitemOpen
  \bibfield  {author} {\bibinfo {author} {\bibfnamefont {A.}~\bibnamefont
  {Ohtomo}}\ and\ \bibinfo {author} {\bibfnamefont {H.~Y.}\ \bibnamefont
  {Hwang}},\ }\bibfield  {title} {\bibinfo {title} {A high-mobility electron
  gas at the LaAlO$_3$/SrTiO$_3$ heterointerface},\ }\href
  {https://doi.org/10.1038/nature02308} {\bibfield  {journal} {\bibinfo
  {journal} {Nature}\ }\textbf {\bibinfo {volume} {427}},\ \bibinfo {pages}
  {423} (\bibinfo {year} {2004})}\BibitemShut {NoStop}%
\bibitem [{\citenamefont {Brinkman}\ \emph {et~al.}(2007)\citenamefont
  {Brinkman}, \citenamefont {Huijben}, \citenamefont {van Zalk}, \citenamefont
  {Huijben}, \citenamefont {Zeitler}, \citenamefont {Maan}, \citenamefont
  {van~der Wiel}, \citenamefont {Rijnders}, \citenamefont {Blank},\ and\
  \citenamefont {Hilgenkamp}}]{NatureMater.6.493}%
  \BibitemOpen
  \bibfield  {author} {\bibinfo {author} {\bibfnamefont {A.}~\bibnamefont
  {Brinkman}}, \bibinfo {author} {\bibfnamefont {M.}~\bibnamefont {Huijben}},
  \bibinfo {author} {\bibfnamefont {M.}~\bibnamefont {van Zalk}}, \bibinfo
  {author} {\bibfnamefont {J.}~\bibnamefont {Huijben}}, \bibinfo {author}
  {\bibfnamefont {U.}~\bibnamefont {Zeitler}}, \bibinfo {author} {\bibfnamefont
  {J.~C.}\ \bibnamefont {Maan}}, \bibinfo {author} {\bibfnamefont {W.~G.}\
  \bibnamefont {van~der Wiel}}, \bibinfo {author} {\bibfnamefont
  {G.}~\bibnamefont {Rijnders}}, \bibinfo {author} {\bibfnamefont {D.~H.~A.}\
  \bibnamefont {Blank}},\ and\ \bibinfo {author} {\bibfnamefont
  {H.}~\bibnamefont {Hilgenkamp}},\ }\bibfield  {title} {\bibinfo {title}
  {Magnetic effects at the interface between non-magnetic oxides},\ }\href
  {https://doi.org/10.1038/nmat1931} {\bibfield  {journal} {\bibinfo  {journal}
  {Nature Mater.}\ }\textbf {\bibinfo {volume} {6}},\ \bibinfo {pages} {493}
  (\bibinfo {year} {2007})}\BibitemShut {NoStop}%
\bibitem [{\citenamefont {Reyren}\ \emph {et~al.}(2007)\citenamefont {Reyren},
  \citenamefont {Thiel}, \citenamefont {Caviglia}, \citenamefont {Kourkoutis},
  \citenamefont {Hammerl}, \citenamefont {Richter}, \citenamefont {Schneider},
  \citenamefont {Kopp}, \citenamefont {R\"uetschi}, \citenamefont {Jaccard},
  \citenamefont {Gabay}, \citenamefont {Muller}, \citenamefont {Triscone},\
  and\ \citenamefont {Mannhart}}]{Science.317.1196}%
  \BibitemOpen
  \bibfield  {author} {\bibinfo {author} {\bibfnamefont {N.}~\bibnamefont
  {Reyren}}, \bibinfo {author} {\bibfnamefont {S.}~\bibnamefont {Thiel}},
  \bibinfo {author} {\bibfnamefont {A.~D.}\ \bibnamefont {Caviglia}}, \bibinfo
  {author} {\bibfnamefont {L.~F.}\ \bibnamefont {Kourkoutis}}, \bibinfo
  {author} {\bibfnamefont {G.}~\bibnamefont {Hammerl}}, \bibinfo {author}
  {\bibfnamefont {C.}~\bibnamefont {Richter}}, \bibinfo {author} {\bibfnamefont
  {C.~W.}\ \bibnamefont {Schneider}}, \bibinfo {author} {\bibfnamefont
  {T.}~\bibnamefont {Kopp}}, \bibinfo {author} {\bibfnamefont {A.-S.}\
  \bibnamefont {R\"uetschi}}, \bibinfo {author} {\bibfnamefont
  {D.}~\bibnamefont {Jaccard}}, \bibinfo {author} {\bibfnamefont
  {M.}~\bibnamefont {Gabay}}, \bibinfo {author} {\bibfnamefont {D.~A.}\
  \bibnamefont {Muller}}, \bibinfo {author} {\bibfnamefont {J.-M.}\
  \bibnamefont {Triscone}},\ and\ \bibinfo {author} {\bibfnamefont
  {J.}~\bibnamefont {Mannhart}},\ }\bibfield  {title} {\bibinfo {title}
  {Superconducting interfaces between insulating oxides},\ }\href
  {https://doi.org/10.1126/science.1146006} {\bibfield  {journal} {\bibinfo
  {journal} {Science}\ }\textbf {\bibinfo {volume} {317}},\ \bibinfo {pages}
  {1196} (\bibinfo {year} {2007})}\BibitemShut {NoStop}%
\bibitem [{\citenamefont {Caviglia}\ \emph {et~al.}(2008)\citenamefont
  {Caviglia}, \citenamefont {Gariglio}, \citenamefont {Reyren}, \citenamefont
  {Jaccard}, \citenamefont {Schneider}, \citenamefont {Gabay}, \citenamefont
  {Thiel}, \citenamefont {Hammerl}, \citenamefont {Mannhart},\ and\
  \citenamefont {Triscone}}]{Nature.456.624}%
  \BibitemOpen
  \bibfield  {author} {\bibinfo {author} {\bibfnamefont {A.~D.}\ \bibnamefont
  {Caviglia}}, \bibinfo {author} {\bibfnamefont {S.}~\bibnamefont {Gariglio}},
  \bibinfo {author} {\bibfnamefont {N.}~\bibnamefont {Reyren}}, \bibinfo
  {author} {\bibfnamefont {D.}~\bibnamefont {Jaccard}}, \bibinfo {author}
  {\bibfnamefont {T.}~\bibnamefont {Schneider}}, \bibinfo {author}
  {\bibfnamefont {M.}~\bibnamefont {Gabay}}, \bibinfo {author} {\bibfnamefont
  {S.}~\bibnamefont {Thiel}}, \bibinfo {author} {\bibfnamefont
  {G.}~\bibnamefont {Hammerl}}, \bibinfo {author} {\bibfnamefont
  {J.}~\bibnamefont {Mannhart}},\ and\ \bibinfo {author} {\bibfnamefont
  {J.-M.}\ \bibnamefont {Triscone}},\ }\bibfield  {title} {\bibinfo {title}
  {Electric field control of the LaAlO$_3$/SrTiO$_3$ interface ground state},\
  }\href {https://doi.org/10.1038/nature07576} {\bibfield  {journal} {\bibinfo
  {journal} {Nature}\ }\textbf {\bibinfo {volume} {456}},\ \bibinfo {pages}
  {624} (\bibinfo {year} {2008})}\BibitemShut {NoStop}%
\bibitem [{\citenamefont {Yang}\ \emph {et~al.}(2016)\citenamefont {Yang},
  \citenamefont {Nazir}, \citenamefont {Behtash},\ and\ \citenamefont
  {Cheng}}]{SciRep.6.34667}%
  \BibitemOpen
  \bibfield  {author} {\bibinfo {author} {\bibfnamefont {K.}~\bibnamefont
  {Yang}}, \bibinfo {author} {\bibfnamefont {S.}~\bibnamefont {Nazir}},
  \bibinfo {author} {\bibfnamefont {M.}~\bibnamefont {Behtash}},\ and\ \bibinfo
  {author} {\bibfnamefont {J.}~\bibnamefont {Cheng}},\ }\bibfield  {title}
  {\bibinfo {title} {High-throughput design of two-dimensional electron gas
  systems based on polar/nonpolar perovskite oxide heterostructures},\ }\href
  {https://doi.org/10.1038/srep34667} {\bibfield  {journal} {\bibinfo
  {journal} {Sci. Rep.}\ }\textbf {\bibinfo {volume} {6}},\ \bibinfo {pages}
  {34667} (\bibinfo {year} {2016})}\BibitemShut {NoStop}%
\bibitem [{\citenamefont {Yin}\ \emph {et~al.}(2015)\citenamefont {Yin},
  \citenamefont {Aguado-Puente}, \citenamefont {Qu},\ and\ \citenamefont
  {Artacho}}]{PhysRevB.92.115406}%
  \BibitemOpen
  \bibfield  {author} {\bibinfo {author} {\bibfnamefont {B.}~\bibnamefont
  {Yin}}, \bibinfo {author} {\bibfnamefont {P.}~\bibnamefont {Aguado-Puente}},
  \bibinfo {author} {\bibfnamefont {S.}~\bibnamefont {Qu}},\ and\ \bibinfo
  {author} {\bibfnamefont {E.}~\bibnamefont {Artacho}},\ }\bibfield  {title}
  {\bibinfo {title} {Two-dimensional electron gas at the PbTiO$_3$/SrTiO$_3$
  interface: An ab initio study},\ }\href
  {https://doi.org/10.1103/PhysRevB.92.115406} {\bibfield  {journal} {\bibinfo
  {journal} {Phys. Rev. B}\ }\textbf {\bibinfo {volume} {92}},\ \bibinfo
  {pages} {115406} (\bibinfo {year} {2015})}\BibitemShut {NoStop}%
\bibitem [{\citenamefont {Ruan}\ \emph {et~al.}(2015)\citenamefont {Ruan},
  \citenamefont {Qiu}, \citenamefont {Yuan}, \citenamefont {Ji}, \citenamefont
  {Wang}, \citenamefont {Li},\ and\ \citenamefont
  {Wu}}]{ApplPhysLett.107.232902}%
  \BibitemOpen
  \bibfield  {author} {\bibinfo {author} {\bibfnamefont {J.}~\bibnamefont
  {Ruan}}, \bibinfo {author} {\bibfnamefont {X.}~\bibnamefont {Qiu}}, \bibinfo
  {author} {\bibfnamefont {Z.}~\bibnamefont {Yuan}}, \bibinfo {author}
  {\bibfnamefont {D.}~\bibnamefont {Ji}}, \bibinfo {author} {\bibfnamefont
  {P.}~\bibnamefont {Wang}}, \bibinfo {author} {\bibfnamefont {A.}~\bibnamefont
  {Li}},\ and\ \bibinfo {author} {\bibfnamefont {D.}~\bibnamefont {Wu}},\
  }\bibfield  {title} {\bibinfo {title} {Improved memory functions in
  multiferroic tunnel junctions with a dielectric/ferroelectric composite
  barrier},\ }\href {https://doi.org/10.1063/1.4937390} {\bibfield  {journal}
  {\bibinfo  {journal} {Appl. Phys. Lett.}\ }\textbf {\bibinfo {volume}
  {107}},\ \bibinfo {pages} {232902} (\bibinfo {year} {2015})}\BibitemShut
  {NoStop}%
\bibitem [{\citenamefont {Wu}\ \emph {et~al.}(2016)\citenamefont {Wu},
  \citenamefont {Shen}, \citenamefont {Yang}, \citenamefont {Zhou},
  \citenamefont {Chen},\ and\ \citenamefont {Feng}}]{PhysRevB.94.155420}%
  \BibitemOpen
  \bibfield  {author} {\bibinfo {author} {\bibfnamefont {Q.}~\bibnamefont
  {Wu}}, \bibinfo {author} {\bibfnamefont {L.}~\bibnamefont {Shen}}, \bibinfo
  {author} {\bibfnamefont {M.}~\bibnamefont {Yang}}, \bibinfo {author}
  {\bibfnamefont {J.}~\bibnamefont {Zhou}}, \bibinfo {author} {\bibfnamefont
  {J.}~\bibnamefont {Chen}},\ and\ \bibinfo {author} {\bibfnamefont {Y.~P.}\
  \bibnamefont {Feng}},\ }\bibfield  {title} {\bibinfo {title} {Giant tunneling
  electroresistance induced by ferroelectrically switchable two-dimensional
  electron gas at nonpolar BaTiO$_3$/SrTiO$_3$ interface},\ }\href
  {https://doi.org/10.1103/PhysRevB.94.155420} {\bibfield  {journal} {\bibinfo
  {journal} {Phys. Rev. B}\ }\textbf {\bibinfo {volume} {94}},\ \bibinfo
  {pages} {155420} (\bibinfo {year} {2016})}\BibitemShut {NoStop}%
\bibitem [{\citenamefont {Tsymbal}\ and\ \citenamefont
  {Kohlstedt}(2006)}]{Science.313.181}%
  \BibitemOpen
  \bibfield  {author} {\bibinfo {author} {\bibfnamefont {E.~Y.}\ \bibnamefont
  {Tsymbal}}\ and\ \bibinfo {author} {\bibfnamefont {H.}~\bibnamefont
  {Kohlstedt}},\ }\bibfield  {title} {\bibinfo {title} {Tunneling across a
  ferroelectric},\ }\href {https://doi.org/10.1126/science.1126230} {\bibfield
  {journal} {\bibinfo  {journal} {Science}\ }\textbf {\bibinfo {volume}
  {313}},\ \bibinfo {pages} {181} (\bibinfo {year} {2006})}\BibitemShut
  {NoStop}%
\bibitem [{\citenamefont {Velev}\ \emph {et~al.}(2009)\citenamefont {Velev},
  \citenamefont {Duan}, \citenamefont {Burton}, \citenamefont {Smogunov},
  \citenamefont {Niranjan}, \citenamefont {Tosatti}, \citenamefont {Jaswal},\
  and\ \citenamefont {Tsymbal}}]{NanoLett.9.427}%
  \BibitemOpen
  \bibfield  {author} {\bibinfo {author} {\bibfnamefont {J.~P.}\ \bibnamefont
  {Velev}}, \bibinfo {author} {\bibfnamefont {C.-G.}\ \bibnamefont {Duan}},
  \bibinfo {author} {\bibfnamefont {J.~D.}\ \bibnamefont {Burton}}, \bibinfo
  {author} {\bibfnamefont {A.}~\bibnamefont {Smogunov}}, \bibinfo {author}
  {\bibfnamefont {M.~K.}\ \bibnamefont {Niranjan}}, \bibinfo {author}
  {\bibfnamefont {E.}~\bibnamefont {Tosatti}}, \bibinfo {author} {\bibfnamefont
  {S.~S.}\ \bibnamefont {Jaswal}},\ and\ \bibinfo {author} {\bibfnamefont
  {E.~Y.}\ \bibnamefont {Tsymbal}},\ }\bibfield  {title} {\bibinfo {title}
  {Magnetic tunnel junctions with ferroelectric barriers: Prediction of four
  resistance states from first principles},\ }\href
  {https://doi.org/10.1021/nl803318d} {\bibfield  {journal} {\bibinfo
  {journal} {Nano Lett.}\ }\textbf {\bibinfo {volume} {9}},\ \bibinfo {pages}
  {427} (\bibinfo {year} {2009})}\BibitemShut {NoStop}%
\bibitem [{\citenamefont {Wang}\ \emph {et~al.}(2019)\citenamefont {Wang},
  \citenamefont {Li}, \citenamefont {Li}, \citenamefont {Zhang}, \citenamefont
  {Wang}, \citenamefont {Chen}, \citenamefont {Liu}, \citenamefont {Wang},
  \citenamefont {Zhao},\ and\ \citenamefont {Mei}}]{ChinPhysB.28.047101}%
  \BibitemOpen
  \bibfield  {author} {\bibinfo {author} {\bibfnamefont {F.-N.}\ \bibnamefont
  {Wang}}, \bibinfo {author} {\bibfnamefont {J.-C.}\ \bibnamefont {Li}},
  \bibinfo {author} {\bibfnamefont {Y.}~\bibnamefont {Li}}, \bibinfo {author}
  {\bibfnamefont {X.-M.}\ \bibnamefont {Zhang}}, \bibinfo {author}
  {\bibfnamefont {X.-J.}\ \bibnamefont {Wang}}, \bibinfo {author}
  {\bibfnamefont {Y.-F.}\ \bibnamefont {Chen}}, \bibinfo {author}
  {\bibfnamefont {J.}~\bibnamefont {Liu}}, \bibinfo {author} {\bibfnamefont
  {C.-L.}\ \bibnamefont {Wang}}, \bibinfo {author} {\bibfnamefont {M.-L.}\
  \bibnamefont {Zhao}},\ and\ \bibinfo {author} {\bibfnamefont {L.-M.}\
  \bibnamefont {Mei}},\ }\bibfield  {title} {\bibinfo {title} {Prediction of
  high-mobility two-dimensional electron gas at KTaO$_3$-based
  heterointerfaces},\ }\href {https://doi.org/10.1088/1674-1056/28/4/047101}
  {\bibfield  {journal} {\bibinfo  {journal} {Chin. Phys. B}\ }\textbf
  {\bibinfo {volume} {28}},\ \bibinfo {pages} {047101} (\bibinfo {year}
  {2019})}\BibitemShut {NoStop}%
\bibitem [{\citenamefont {Murray}\ and\ \citenamefont
  {Vanderbilt}(2009)}]{PhysRevB.79.100102}%
  \BibitemOpen
  \bibfield  {author} {\bibinfo {author} {\bibfnamefont {{\'E}.~D.}\
  \bibnamefont {Murray}}\ and\ \bibinfo {author} {\bibfnamefont
  {D.}~\bibnamefont {Vanderbilt}},\ }\bibfield  {title} {\bibinfo {title}
  {Theoretical investigation of polarization-compensated II-IV/I-V perovskite
  superlattices},\ }\href {https://doi.org/10.1103/PhysRevB.79.100102}
  {\bibfield  {journal} {\bibinfo  {journal} {Phys. Rev. B}\ }\textbf {\bibinfo
  {volume} {79}},\ \bibinfo {pages} {100102} (\bibinfo {year}
  {2009})}\BibitemShut {NoStop}%
\bibitem [{\citenamefont {Niranjan}\ \emph {et~al.}(2009)\citenamefont
  {Niranjan}, \citenamefont {Wang}, \citenamefont {Jaswal},\ and\ \citenamefont
  {Tsymbal}}]{PhysRevLett.103.016804}%
  \BibitemOpen
  \bibfield  {author} {\bibinfo {author} {\bibfnamefont {M.~K.}\ \bibnamefont
  {Niranjan}}, \bibinfo {author} {\bibfnamefont {Y.}~\bibnamefont {Wang}},
  \bibinfo {author} {\bibfnamefont {S.~S.}\ \bibnamefont {Jaswal}},\ and\
  \bibinfo {author} {\bibfnamefont {E.~Y.}\ \bibnamefont {Tsymbal}},\
  }\bibfield  {title} {\bibinfo {title} {Prediction of a switchable
  two-dimensional electron gas at ferroelectric oxide interfaces},\ }\href
  {https://doi.org/10.1103/PhysRevLett.103.016804} {\bibfield  {journal}
  {\bibinfo  {journal} {Phys. Rev. Lett.}\ }\textbf {\bibinfo {volume} {103}},\
  \bibinfo {pages} {016804} (\bibinfo {year} {2009})}\BibitemShut {NoStop}%
\bibitem [{\citenamefont {Bristowe}\ \emph {et~al.}(2009)\citenamefont
  {Bristowe}, \citenamefont {Artacho},\ and\ \citenamefont
  {Littlewood}}]{PhysRevB.80.045425}%
  \BibitemOpen
  \bibfield  {author} {\bibinfo {author} {\bibfnamefont {N.~C.}\ \bibnamefont
  {Bristowe}}, \bibinfo {author} {\bibfnamefont {E.}~\bibnamefont {Artacho}},\
  and\ \bibinfo {author} {\bibfnamefont {P.~B.}\ \bibnamefont {Littlewood}},\
  }\bibfield  {title} {\bibinfo {title} {Oxide superlattices with alternating
  $p$ and $n$ interfaces},\ }\href {https://doi.org/10.1103/PhysRevB.80.045425}
  {\bibfield  {journal} {\bibinfo  {journal} {Phys. Rev. B}\ }\textbf {\bibinfo
  {volume} {80}},\ \bibinfo {pages} {045425} (\bibinfo {year}
  {2009})}\BibitemShut {NoStop}%
\bibitem [{\citenamefont {Wang}\ \emph {et~al.}(2009)\citenamefont {Wang},
  \citenamefont {Niranjan}, \citenamefont {Jaswal},\ and\ \citenamefont
  {Tsymbal}}]{PhysRevB.80.165130}%
  \BibitemOpen
  \bibfield  {author} {\bibinfo {author} {\bibfnamefont {Y.}~\bibnamefont
  {Wang}}, \bibinfo {author} {\bibfnamefont {M.~K.}\ \bibnamefont {Niranjan}},
  \bibinfo {author} {\bibfnamefont {S.~S.}\ \bibnamefont {Jaswal}},\ and\
  \bibinfo {author} {\bibfnamefont {E.~Y.}\ \bibnamefont {Tsymbal}},\
  }\bibfield  {title} {\bibinfo {title} {First-principles studies of a
  two-dimensional electron gas at the interface in ferroelectric oxide
  heterostructures},\ }\href {https://doi.org/10.1103/PhysRevB.80.165130}
  {\bibfield  {journal} {\bibinfo  {journal} {Phys. Rev. B}\ }\textbf {\bibinfo
  {volume} {80}},\ \bibinfo {pages} {165130} (\bibinfo {year}
  {2009})}\BibitemShut {NoStop}%
\bibitem [{\citenamefont {Stengel}\ and\ \citenamefont
  {Vanderbilt}(2009)}]{PhysRevB.80.241103}%
  \BibitemOpen
  \bibfield  {author} {\bibinfo {author} {\bibfnamefont {M.}~\bibnamefont
  {Stengel}}\ and\ \bibinfo {author} {\bibfnamefont {D.}~\bibnamefont
  {Vanderbilt}},\ }\bibfield  {title} {\bibinfo {title} {Berry-phase theory of
  polar discontinuities at oxide-oxide interfaces},\ }\href
  {https://doi.org/10.1103/PhysRevB.80.241103} {\bibfield  {journal} {\bibinfo
  {journal} {Phys. Rev. B}\ }\textbf {\bibinfo {volume} {80}},\ \bibinfo
  {pages} {241103} (\bibinfo {year} {2009})}\BibitemShut {NoStop}%
\bibitem [{\citenamefont {Das}\ \emph {et~al.}(2010)\citenamefont {Das},
  \citenamefont {Spaldin}, \citenamefont {Waghmare},\ and\ \citenamefont
  {Saha-Dasgupta}}]{PhysRevB.81.235112}%
  \BibitemOpen
  \bibfield  {author} {\bibinfo {author} {\bibfnamefont {H.}~\bibnamefont
  {Das}}, \bibinfo {author} {\bibfnamefont {N.~A.}\ \bibnamefont {Spaldin}},
  \bibinfo {author} {\bibfnamefont {U.~V.}\ \bibnamefont {Waghmare}},\ and\
  \bibinfo {author} {\bibfnamefont {T.}~\bibnamefont {Saha-Dasgupta}},\
  }\bibfield  {title} {\bibinfo {title} {Chemical control of polar behavior in
  bicomponent short-period superlattices},\ }\href
  {https://doi.org/10.1103/PhysRevB.81.235112} {\bibfield  {journal} {\bibinfo
  {journal} {Phys. Rev. B}\ }\textbf {\bibinfo {volume} {81}},\ \bibinfo
  {pages} {235112} (\bibinfo {year} {2010})}\BibitemShut {NoStop}%
\bibitem [{\citenamefont {Cooper}(2012)}]{PhysRevB.85.235109}%
  \BibitemOpen
  \bibfield  {author} {\bibinfo {author} {\bibfnamefont {V.~R.}\ \bibnamefont
  {Cooper}},\ }\bibfield  {title} {\bibinfo {title} {Enhanced carrier
  mobilities in two-dimensional electron gases at III-III/I-V oxide
  heterostructure interfaces},\ }\href
  {https://doi.org/10.1103/PhysRevB.85.235109} {\bibfield  {journal} {\bibinfo
  {journal} {Phys. Rev. B}\ }\textbf {\bibinfo {volume} {85}},\ \bibinfo
  {pages} {235109} (\bibinfo {year} {2012})}\BibitemShut {NoStop}%
\bibitem [{\citenamefont {Garc\'{\i}a-Fern\'andez}\ \emph
  {et~al.}(2013)\citenamefont {Garc\'{\i}a-Fern\'andez}, \citenamefont
  {Aguado-Puente},\ and\ \citenamefont {Junquera}}]{PhysRevB.87.085305}%
  \BibitemOpen
  \bibfield  {author} {\bibinfo {author} {\bibfnamefont {P.}~\bibnamefont
  {Garc\'{\i}a-Fern\'andez}}, \bibinfo {author} {\bibfnamefont
  {P.}~\bibnamefont {Aguado-Puente}},\ and\ \bibinfo {author} {\bibfnamefont
  {J.}~\bibnamefont {Junquera}},\ }\bibfield  {title} {\bibinfo {title}
  {Lattice screening of the polar catastrophe and hidden in-plane polarization
  in KNbO$_3$/BaTiO$_3$ interfaces},\ }\href
  {https://doi.org/10.1103/PhysRevB.87.085305} {\bibfield  {journal} {\bibinfo
  {journal} {Phys. Rev. B}\ }\textbf {\bibinfo {volume} {87}},\ \bibinfo
  {pages} {085305} (\bibinfo {year} {2013})}\BibitemShut {NoStop}%
\bibitem [{\citenamefont {Chen}\ \emph {et~al.}(2018)\citenamefont {Chen},
  \citenamefont {Fang}, \citenamefont {Zhang}, \citenamefont {Zhao},\ and\
  \citenamefont {Ren}}]{SciRep.8.467}%
  \BibitemOpen
  \bibfield  {author} {\bibinfo {author} {\bibfnamefont {C.}~\bibnamefont
  {Chen}}, \bibinfo {author} {\bibfnamefont {L.}~\bibnamefont {Fang}}, \bibinfo
  {author} {\bibfnamefont {J.}~\bibnamefont {Zhang}}, \bibinfo {author}
  {\bibfnamefont {G.}~\bibnamefont {Zhao}},\ and\ \bibinfo {author}
  {\bibfnamefont {W.}~\bibnamefont {Ren}},\ }\bibfield  {title} {\bibinfo
  {title} {Thickness control of the spin-polarized two-dimensional electron gas
  in LaAlO$_3$/BaTiO$_3$ superlattices},\ }\href
  {https://doi.org/10.1038/s41598-017-18858-x} {\bibfield  {journal} {\bibinfo
  {journal} {Sci. Rep.}\ }\textbf {\bibinfo {volume} {8}},\ \bibinfo {pages}
  {467} (\bibinfo {year} {2018})}\BibitemShut {NoStop}%
\bibitem [{\citenamefont {Li}\ \emph {et~al.}(2019)\citenamefont {Li},
  \citenamefont {Huang}, \citenamefont {Peng}, \citenamefont {Xiong},
  \citenamefont {Xiao}, \citenamefont {Yan}, \citenamefont {Cao}, \citenamefont
  {Tang},\ and\ \citenamefont {Li}}]{RCSAdv.9.35499}%
  \BibitemOpen
  \bibfield  {author} {\bibinfo {author} {\bibfnamefont {G.}~\bibnamefont
  {Li}}, \bibinfo {author} {\bibfnamefont {H.}~\bibnamefont {Huang}}, \bibinfo
  {author} {\bibfnamefont {S.}~\bibnamefont {Peng}}, \bibinfo {author}
  {\bibfnamefont {Y.}~\bibnamefont {Xiong}}, \bibinfo {author} {\bibfnamefont
  {Y.}~\bibnamefont {Xiao}}, \bibinfo {author} {\bibfnamefont {S.}~\bibnamefont
  {Yan}}, \bibinfo {author} {\bibfnamefont {Y.}~\bibnamefont {Cao}}, \bibinfo
  {author} {\bibfnamefont {M.}~\bibnamefont {Tang}},\ and\ \bibinfo {author}
  {\bibfnamefont {Z.}~\bibnamefont {Li}},\ }\bibfield  {title} {\bibinfo
  {title} {Two-dimensional polar metals in KNbO$_3$/BaTiO$_3$ superlattices:
  first-principle calculations},\ }\href {https://doi.org/10.1039/C9RA06209B}
  {\bibfield  {journal} {\bibinfo  {journal} {RSC Adv.}\ }\textbf {\bibinfo
  {volume} {9}},\ \bibinfo {pages} {35499} (\bibinfo {year}
  {2019})}\BibitemShut {NoStop}%
\bibitem [{\citenamefont {Fang}\ \emph {et~al.}(2019)\citenamefont {Fang},
  \citenamefont {Chen}, \citenamefont {Yang}, \citenamefont {Wu}, \citenamefont
  {Hu}, \citenamefont {Zhao}, \citenamefont {Zhu},\ and\ \citenamefont
  {Ren}}]{PhysChemChemPhys.21.8046}%
  \BibitemOpen
  \bibfield  {author} {\bibinfo {author} {\bibfnamefont {L.}~\bibnamefont
  {Fang}}, \bibinfo {author} {\bibfnamefont {C.}~\bibnamefont {Chen}}, \bibinfo
  {author} {\bibfnamefont {Y.}~\bibnamefont {Yang}}, \bibinfo {author}
  {\bibfnamefont {Y.}~\bibnamefont {Wu}}, \bibinfo {author} {\bibfnamefont
  {T.}~\bibnamefont {Hu}}, \bibinfo {author} {\bibfnamefont {G.}~\bibnamefont
  {Zhao}}, \bibinfo {author} {\bibfnamefont {Q.}~\bibnamefont {Zhu}},\ and\
  \bibinfo {author} {\bibfnamefont {W.}~\bibnamefont {Ren}},\ }\bibfield
  {title} {\bibinfo {title} {First-principles studies of a two-dimensional
  electron gas at the interface of polar/polar LaAlO$_3$/KNbO$_3$
  superlattices},\ }\href {https://doi.org/10.1039/C8CP07202G} {\bibfield
  {journal} {\bibinfo  {journal} {Phys. Chem. Chem. Phys.}\ }\textbf {\bibinfo
  {volume} {21}},\ \bibinfo {pages} {8046} (\bibinfo {year}
  {2019})}\BibitemShut {NoStop}%
\bibitem [{\citenamefont {Das}\ \emph {et~al.}(2011)\citenamefont {Das},
  \citenamefont {Waghmare},\ and\ \citenamefont
  {Saha-Dasgupta}}]{JApplPhys.109.066107}%
  \BibitemOpen
  \bibfield  {author} {\bibinfo {author} {\bibfnamefont {H.}~\bibnamefont
  {Das}}, \bibinfo {author} {\bibfnamefont {U.~V.}\ \bibnamefont {Waghmare}},\
  and\ \bibinfo {author} {\bibfnamefont {T.}~\bibnamefont {Saha-Dasgupta}},\
  }\bibfield  {title} {\bibinfo {title} {Piezoelectrics by design: A route
  through short-period perovskite superlattices},\ }\href
  {https://doi.org/10.1063/1.3561843} {\bibfield  {journal} {\bibinfo
  {journal} {J. Appl. Phys.}\ }\textbf {\bibinfo {volume} {109}},\ \bibinfo
  {pages} {066107} (\bibinfo {year} {2011})}\BibitemShut {NoStop}%
\bibitem [{\citenamefont {Zhu}(2018)}]{ChinPhysB.27.027701}%
  \BibitemOpen
  \bibfield  {author} {\bibinfo {author} {\bibfnamefont {Z.}~\bibnamefont
  {Zhu}},\ }\bibfield  {title} {\bibinfo {title} {First-principles study of
  polarization and piezoelectricity behavior in tetragonal PbTiO$_3$-based
  superlattices},\ }\href {https://doi.org/10.1088/1674-1056/27/2/027701}
  {\bibfield  {journal} {\bibinfo  {journal} {Chin. Phys. B}\ }\textbf
  {\bibinfo {volume} {27}},\ \bibinfo {pages} {027701} (\bibinfo {year}
  {2018})}\BibitemShut {NoStop}%
\bibitem [{\citenamefont {Park}\ and\ \citenamefont
  {Shrout}(1997)}]{JApplPhys.82.1804}%
  \BibitemOpen
  \bibfield  {author} {\bibinfo {author} {\bibfnamefont {S.-E.}\ \bibnamefont
  {Park}}\ and\ \bibinfo {author} {\bibfnamefont {T.~R.}\ \bibnamefont
  {Shrout}},\ }\bibfield  {title} {\bibinfo {title} {Ultrahigh strain and
  piezoelectric behavior in relaxor based ferroelectric single crystals},\
  }\href {https://doi.org/10.1063/1.365983} {\bibfield  {journal} {\bibinfo
  {journal} {J. Appl. Phys.}\ }\textbf {\bibinfo {volume} {82}},\ \bibinfo
  {pages} {1804} (\bibinfo {year} {1997})}\BibitemShut {NoStop}%
\bibitem [{\citenamefont {Guo}\ \emph {et~al.}(2000)\citenamefont {Guo},
  \citenamefont {Cross}, \citenamefont {Park}, \citenamefont {Noheda},
  \citenamefont {Cox},\ and\ \citenamefont {Shirane}}]{PhysRevLett.84.5423}%
  \BibitemOpen
  \bibfield  {author} {\bibinfo {author} {\bibfnamefont {R.}~\bibnamefont
  {Guo}}, \bibinfo {author} {\bibfnamefont {L.~E.}\ \bibnamefont {Cross}},
  \bibinfo {author} {\bibfnamefont {S.-E.}\ \bibnamefont {Park}}, \bibinfo
  {author} {\bibfnamefont {B.}~\bibnamefont {Noheda}}, \bibinfo {author}
  {\bibfnamefont {D.~E.}\ \bibnamefont {Cox}},\ and\ \bibinfo {author}
  {\bibfnamefont {G.}~\bibnamefont {Shirane}},\ }\bibfield  {title} {\bibinfo
  {title} {Origin of the high piezoelectric response in
  PbZr$_{1-x}$Ti$_x$O$_3$},\ }\href
  {https://doi.org/10.1103/PhysRevLett.84.5423} {\bibfield  {journal} {\bibinfo
   {journal} {Phys. Rev. Lett.}\ }\textbf {\bibinfo {volume} {84}},\ \bibinfo
  {pages} {5423} (\bibinfo {year} {2000})}\BibitemShut {NoStop}%
\bibitem [{\citenamefont {Gonze}\ \emph {et~al.}(2002)\citenamefont {Gonze},
  \citenamefont {Beuken}, \citenamefont {Caracas}, \citenamefont {Detraux},
  \citenamefont {Fuchs}, \citenamefont {Rignanese}, \citenamefont {Sindic},
  \citenamefont {Verstraete}, \citenamefont {Zerah}, \citenamefont {Jollet},
  \citenamefont {Torrent}, \citenamefont {Roy}, \citenamefont {Mikami},
  \citenamefont {Ghosez}, \citenamefont {Raty},\ and\ \citenamefont
  {Allan}}]{abinit}%
  \BibitemOpen
  \bibfield  {author} {\bibinfo {author} {\bibfnamefont {X.}~\bibnamefont
  {Gonze}}, \bibinfo {author} {\bibfnamefont {J.-M.}\ \bibnamefont {Beuken}},
  \bibinfo {author} {\bibfnamefont {R.}~\bibnamefont {Caracas}}, \bibinfo
  {author} {\bibfnamefont {F.}~\bibnamefont {Detraux}}, \bibinfo {author}
  {\bibfnamefont {M.}~\bibnamefont {Fuchs}}, \bibinfo {author} {\bibfnamefont
  {G.-M.}\ \bibnamefont {Rignanese}}, \bibinfo {author} {\bibfnamefont
  {L.}~\bibnamefont {Sindic}}, \bibinfo {author} {\bibfnamefont
  {M.}~\bibnamefont {Verstraete}}, \bibinfo {author} {\bibfnamefont
  {G.}~\bibnamefont {Zerah}}, \bibinfo {author} {\bibfnamefont
  {F.}~\bibnamefont {Jollet}}, \bibinfo {author} {\bibfnamefont
  {M.}~\bibnamefont {Torrent}}, \bibinfo {author} {\bibfnamefont
  {A.}~\bibnamefont {Roy}}, \bibinfo {author} {\bibfnamefont {M.}~\bibnamefont
  {Mikami}}, \bibinfo {author} {\bibfnamefont {P.}~\bibnamefont {Ghosez}},
  \bibinfo {author} {\bibfnamefont {J.-Y.}\ \bibnamefont {Raty}},\ and\
  \bibinfo {author} {\bibfnamefont {D.~C.}\ \bibnamefont {Allan}},\ }\bibfield
  {title} {\bibinfo {title} {First-principles computation of material
  properties: the ABINIT software project},\ }\href
  {https://doi.org/10.1016/S0927-0256(02)0325-7} {\bibfield  {journal}
  {\bibinfo  {journal} {Comput. Mater. Sci.}\ }\textbf {\bibinfo {volume}
  {25}},\ \bibinfo {pages} {478} (\bibinfo {year} {2002})}\BibitemShut
  {NoStop}%
\bibitem [{\citenamefont {Rappe}\ \emph {et~al.}(1990)\citenamefont {Rappe},
  \citenamefont {Rabe}, \citenamefont {Kaxiras},\ and\ \citenamefont
  {Joannopoulos}}]{PhysRevB.41.1227}%
  \BibitemOpen
  \bibfield  {author} {\bibinfo {author} {\bibfnamefont {A.~M.}\ \bibnamefont
  {Rappe}}, \bibinfo {author} {\bibfnamefont {K.~M.}\ \bibnamefont {Rabe}},
  \bibinfo {author} {\bibfnamefont {E.}~\bibnamefont {Kaxiras}},\ and\ \bibinfo
  {author} {\bibfnamefont {J.~D.}\ \bibnamefont {Joannopoulos}},\ }\bibfield
  {title} {\bibinfo {title} {Optimized pseudopotentials},\ }\href
  {https://doi.org/10.1103/PhysRevB.41.1227} {\bibfield  {journal} {\bibinfo
  {journal} {Phys. Rev. B}\ }\textbf {\bibinfo {volume} {41}},\ \bibinfo
  {pages} {1227} (\bibinfo {year} {1990})}\BibitemShut {NoStop}%
\bibitem [{\citenamefont
  {Lebedev}(2009{\natexlab{b}})}]{PhysSolidState.51.362}%
  \BibitemOpen
  \bibfield  {author} {\bibinfo {author} {\bibfnamefont {A.~I.}\ \bibnamefont
  {Lebedev}},\ }\bibfield  {title} {\bibinfo {title} {Ab initio calculations of
  phonon spectra in $A$TiO$_3$ perovskite crystals ($A$ = Ca, Sr, Ba, Ra, Cd, Zn,
  Mg, Ge, Sn, Pb)},\ }\href {https://doi.org/10.1134/S1063783409020279}
  {\bibfield  {journal} {\bibinfo  {journal} {Phys. Solid State}\ }\textbf
  {\bibinfo {volume} {51}},\ \bibinfo {pages} {362} (\bibinfo {year}
  {2009}{\natexlab{b}})}\BibitemShut {NoStop}%
\bibitem [{\citenamefont {Lebedev}(2016)}]{PhysSolidState.58.300}%
  \BibitemOpen
  \bibfield  {author} {\bibinfo {author} {\bibfnamefont {A.~I.}\ \bibnamefont
  {Lebedev}},\ }\bibfield  {title} {\bibinfo {title} {Phase transitions and
  metastable states in stressed SrTiO$_3$ films},\ }\href
  {https://doi.org/10.1134/S1063783416020190} {\bibfield  {journal} {\bibinfo
  {journal} {Phys. Solid State}\ }\textbf {\bibinfo {volume} {58}},\ \bibinfo
  {pages} {300} (\bibinfo {year} {2016})}\BibitemShut {NoStop}%
\bibitem [{\citenamefont {Yu}\ and\ \citenamefont
  {Krakauer}(1995)}]{PhysRevLett.74.4067}%
  \BibitemOpen
  \bibfield  {author} {\bibinfo {author} {\bibfnamefont {R.}~\bibnamefont
  {Yu}}\ and\ \bibinfo {author} {\bibfnamefont {H.}~\bibnamefont {Krakauer}},\
  }\bibfield  {title} {\bibinfo {title} {First-principles determination of
  chain-structure instability in KNbO$_3$},\ }\href
  {https://doi.org/10.1103/PhysRevLett.74.4067} {\bibfield  {journal} {\bibinfo
   {journal} {Phys. Rev. Lett.}\ }\textbf {\bibinfo {volume} {74}},\ \bibinfo
  {pages} {4067} (\bibinfo {year} {1995})}\BibitemShut {NoStop}%
\bibitem [{Note1()}]{Note1}%
  \BibitemOpen
  \bibinfo {note} {When numbering irreducible representations in this paper, we
  follow their classification accepted at Bilbao Crystallographic Server~\cite
  {Bilbao}.}\BibitemShut {Stop}%
\bibitem [{\citenamefont {Pertsev}\ \emph {et~al.}(1998)\citenamefont
  {Pertsev}, \citenamefont {Zembilgotov},\ and\ \citenamefont
  {Tagantsev}}]{PhysRevLett.80.1988}%
  \BibitemOpen
  \bibfield  {author} {\bibinfo {author} {\bibfnamefont {N.~A.}\ \bibnamefont
  {Pertsev}}, \bibinfo {author} {\bibfnamefont {A.~G.}\ \bibnamefont
  {Zembilgotov}},\ and\ \bibinfo {author} {\bibfnamefont {A.~K.}\ \bibnamefont
  {Tagantsev}},\ }\bibfield  {title} {\bibinfo {title} {Effect of mechanical
  boundary conditions on phase diagrams of epitaxial ferroelectric thin
  films},\ }\href {https://doi.org/10.1103/PhysRevLett.80.1988} {\bibfield
  {journal} {\bibinfo  {journal} {Phys. Rev. Lett.}\ }\textbf {\bibinfo
  {volume} {80}},\ \bibinfo {pages} {1988} (\bibinfo {year}
  {1998})}\BibitemShut {NoStop}%
\bibitem [{\citenamefont {Di\'eguez}\ \emph {et~al.}(2004)\citenamefont
  {Di\'eguez}, \citenamefont {Tinte}, \citenamefont {Antons}, \citenamefont
  {Bungaro}, \citenamefont {Neaton}, \citenamefont {Rabe},\ and\ \citenamefont
  {Vanderbilt}}]{PhysRevB.69.212101}%
  \BibitemOpen
  \bibfield  {author} {\bibinfo {author} {\bibfnamefont {O.}~\bibnamefont
  {Di\'eguez}}, \bibinfo {author} {\bibfnamefont {S.}~\bibnamefont {Tinte}},
  \bibinfo {author} {\bibfnamefont {A.}~\bibnamefont {Antons}}, \bibinfo
  {author} {\bibfnamefont {C.}~\bibnamefont {Bungaro}}, \bibinfo {author}
  {\bibfnamefont {J.~B.}\ \bibnamefont {Neaton}}, \bibinfo {author}
  {\bibfnamefont {K.~M.}\ \bibnamefont {Rabe}},\ and\ \bibinfo {author}
  {\bibfnamefont {D.}~\bibnamefont {Vanderbilt}},\ }\bibfield  {title}
  {\bibinfo {title} {Ab initio study of the phase diagram of epitaxial
  BaTiO$_3$},\ }\href {https://doi.org/10.1103/PhysRevB.69.212101} {\bibfield
  {journal} {\bibinfo  {journal} {Phys. Rev. B}\ }\textbf {\bibinfo {volume}
  {69}},\ \bibinfo {pages} {212101} (\bibinfo {year} {2004})}\BibitemShut
  {NoStop}%
\bibitem [{Note2()}]{Note2}%
  \BibitemOpen
  \bibinfo {note} {The periodicity of a superlattice assumes that the electric
  field strengths ${\protect \cal E}_1$ and ${\protect \cal E}_2$ in its layers
  satisfy the condition ${\protect \cal E}_1 x_1 + {\protect \cal E}_2 x_2 =
  0$, where $x_1$ and $x_2$ are the thicknesses of individual layers. This
  condition, combined with the equation $(P_1 + \epsilon _1 {\protect \cal
  E}_1) - (P_2 + \epsilon _2 {\protect \cal E}_2) = \Delta P$, where $P_1$,
  $P_2$, $\epsilon _1$, and $\epsilon _2$ are spontaneous polarizations and
  dielectric constants of two layers, gives a solution to this problem in the
  linear approximation. Solving a nonlinear problem requires more complex
  calculations.}\BibitemShut {Stop}%
\bibitem [{\citenamefont {Resta}\ and\ \citenamefont
  {Vanderbilt}(2007)}]{PhysicsofFerroelectrics2}%
  \BibitemOpen
  \bibfield  {author} {\bibinfo {author} {\bibfnamefont {R.}~\bibnamefont
  {Resta}}\ and\ \bibinfo {author} {\bibfnamefont {D.}~\bibnamefont
  {Vanderbilt}},\ }\bibfield  {title} {\bibinfo {title} {Theory of
  polarization: A modern approach},\ }in\ \href
  {https://doi.org/10.1007/978-3-540-34591-6_2} {\emph {\bibinfo {booktitle}
  {Physics of Ferroelectrics. A Modern Perspective}}},\ \bibinfo {editor}
  {edited by\ \bibinfo {editor} {\bibfnamefont {K.~M.}\ \bibnamefont {Rabe}},
  \bibinfo {editor} {\bibfnamefont {C.~H.}\ \bibnamefont {Ahn}},\ and\ \bibinfo
  {editor} {\bibfnamefont {J.-M.}\ \bibnamefont {Triscone}}}\ (\bibinfo
  {publisher} {Springer-Verlag Berlin Heidelberg},\ \bibinfo {year} {2007})\
  pp.\ \bibinfo {pages} {31--68}\BibitemShut {NoStop}%
\bibitem [{\citenamefont {Zhu}\ \emph {et~al.}(2016)\citenamefont {Zhu},
  \citenamefont {Wang},\ and\ \citenamefont {Fu}}]{ChinPhysLett.33.026302}%
  \BibitemOpen
  \bibfield  {author} {\bibinfo {author} {\bibfnamefont {Z.-Y.}\ \bibnamefont
  {Zhu}}, \bibinfo {author} {\bibfnamefont {S.-Q.}\ \bibnamefont {Wang}},\ and\
  \bibinfo {author} {\bibfnamefont {Y.-M.}\ \bibnamefont {Fu}},\ }\bibfield
  {title} {\bibinfo {title} {First-principles study of properties of strained
  PbTiO$_3$/KTaO$_3$ superlattice},\ }\href
  {https://doi.org/10.1088/0256-307X/33/2/026302} {\bibfield  {journal}
  {\bibinfo  {journal} {Chin. Phys. Lett.}\ }\textbf {\bibinfo {volume} {33}},\
  \bibinfo {pages} {026302} (\bibinfo {year} {2016})}\BibitemShut {NoStop}%
\bibitem [{Bil()}]{Bilbao}%
  \BibitemOpen
  \href {http://www.cryst.ehu.es/} {\bibinfo {title} {Bilbao crystallographic
  server}},\ \bibinfo {note} {\url{http://www.cryst.ehu.es/}}\BibitemShut
  {NoStop}%
\end{thebibliography}
\providecommand{\BIBYu}{Yu}

\end{document}